\documentclass[a4paper]{article}
\usepackage{bbold}
\usepackage[pages=all, color=black, position={current page.south}, placement=bottom, scale=1, opacity=1, vshift=5mm]{background}
\SetBgContents{
 
}      

\usepackage[margin=1in]{geometry} 

\usepackage{amsmath}
\usepackage{amsthm}
\usepackage{amssymb}
\usepackage{multirow}
\usepackage{textcomp}
\usepackage[utf8]{inputenc}
\usepackage{hyperref}
\hypersetup{
	unicode,
}

\usepackage[sort&compress,numbers,square]{natbib}
\theoremstyle{plain}

\theoremstyle{definition}

\usepackage{graphicx, color}
\graphicspath{{fig/}}

\usepackage{algorithm, algpseudocode} 
\usepackage{mathrsfs} 

\usepackage{lipsum}

\usepackage{color}
\title{The adoption of non-pharmaceutical interventions and the role of digital infrastructure during the COVID-19 Pandemic in Colombia, Ecuador, and El Salvador}

\author{Nicol\`o Gozzi$^{1,2}$, Niccol\`o Comini$^{2}$, Nicola Perra$^{1,2}$}

\date{
	$^1$ Networks and Urban Systems Centre, University of Greenwich, UK\\
	$^2$ The World Bank\\
	\vspace{0.5cm}
	\today
}

\begin{document}
	\maketitle

\begin{abstract}
Adherence to the non-pharmaceutical interventions (NPIs) put in place to mitigate the spreading of infectious diseases is a multifaceted problem. Socio-demographic, socio-economic, and epidemiological factors can influence the perceived susceptibility and risk which are known to affect behavior. Furthermore, the adoption of NPIs is dependent upon the barriers, real or perceived, associated with their implementation. We study the determinants of NPIs adherence during the first wave of the COVID-19 Pandemic in Colombia, Ecuador, and El Salvador. Analyses are performed at the level of municipalities and include socio-economic, socio-demographic, and epidemiological indicators. Furthermore, by leveraging a unique dataset comprising tens of millions of internet Speedtest\textsuperscript{\tiny\textregistered} measurements from Ookla\textsuperscript{\tiny\textregistered}, we investigate the quality of the digital infrastructure as a possible barrier to adoption. We use publicly available data provided by Meta capturing aggregated mobility changes as a proxy of adherence to NPIs. Across the three countries considered, we find a significant correlation between mobility drops and digital infrastructure quality. The relationship remains significant after controlling for several factors including socio-economic status, population size, and reported COVID-19 cases. This finding suggests that municipalities with better connectivity were able to afford higher mobility reductions. The link between mobility drops and digital infrastructure quality is stronger at the peak of NPIs stringency. We also find that mobility reductions were more pronounced in larger, denser, and wealthier municipalities. Our work provides new insights on the significance of access to digital tools as an additional factor influencing the ability to follow social distancing guidelines during a health emergency.
\end{abstract}

\maketitle

\section{Introduction}
Stay-at-home mandates, travel bans, face masks, curfews, remote working, and closure of non-essential shops are just some examples of the non-pharmaceutical interventions (NPIs) implemented worldwide to contain the spread of SARS-CoV-2~\cite{perra2020non, Flaxman2020, Snoeijer2021, haug2020ranking, cowling2020impact}.
Though extremely successful public health measures, NPIs are associated with high socio-economic costs, might bring economic activities to a halt, and disrupt social life~\cite{bonaccorsi2020economic, Skarp2021}. 

The literature on the subject suggests that compliance with such measures is a multi-faced problem driven by individual and societal factors. Socio-demographic (e.g., age, gender, educational attainment, country of residence, population density, age pyramid), and epidemiological (e.g., number of cases, deaths, and vaccination rates) indicators play an important role in adherence. Indeed, they modulate the perceived risk, severity, and susceptibility to the threat and can ultimately affect individual behaviors~\cite{hbm1,hbm2,hbm3,perra2020non}. Adherence to NPIs strongly correlates also with socio-economic determinants~\cite{pullano2020population, Duenas2021, Gozzi2021santiago, fraiberger2020uncovering}. Disadvantaged communities, from low to high-income countries, struggled to implement NPIs during the acute phases of the Pandemic~\cite{Gozzi2021santiago, Menaeabg5298, Chang2021, topriceanu2020inequality,  munoz2020racial, yi2020health, mathur2020ethnic}. Undoubtedly, while restrictive measures have disrupted the lives of everyone, the challenges and barriers to adoption faced in implementing them are extremely different across socio-economic strata. Informal jobs and several types of occupations, for example, left fewer possibilities to implement many forms of NPIs~\cite{blundell2020covid, Ahmed2020,perra2020non}. 

Widespread adoption of mobile devices and of the Internet allowed to continue remotely some economic, educational, and social activities~\cite{strusani2020, covid19_digitalinfra}. These new tools, while opening new unprecedented opportunities for many, constituted new barriers for others. Limited and unequal access to digital infrastructures can influence teleworking, e-commerce adoption, distance learning, use of telehealth platforms, and more in general the ability to carry out activities from home thus limiting travels and mobility~\cite{covid19_digitalinfra, work_digitaldivide, Soomro2020, AZUBUIKE2021100022, Eruchalu2021, Watts2020,vakataki2021visualizing}. However, these factors have not been yet extensively explored via a quantitative data-driven approach. Furthermore, the current literature on the subject, apart from a report focused on social distancing in the US during the early phases of 2020 and its relation with access to high-speed internet~\cite{NBERw26982}, is mainly focused on specific contexts such as education and telemedicine probed via surveys~\cite{zhao2020effects,elsalem2020stress,torres2020transition,passanisi2020quarantine,jeste2020changes,Bauer2020OvercomingMH}.

Here, we tackle this limitation and study the adherence to NPIs in the first COVID-19 wave investigating the effects of a wide range of variables, aggregated at the level of municipality, in one lower-middle-income country (El Salvador) and two upper middle-income countries (Colombia and Ecuador) in Latin America. We explore socio-demographic (i.e., population size, population density, the fraction of the population above $60$), socio-economic (i.e., wealth index, GPD per capita), and epidemiological (i.e., number of reported COVID-19 cases) indicators that might affect risk, severity, and susceptibility perception as well as impose barriers to the adoption of NPIs. Additionally, we investigate the quality of the digital infrastructure as a form of barrier that might affect NPIs adherence. To this end, we leverage a unique dataset containing tens of millions of geolocalized internet Speedtest\textsuperscript{\tiny\textregistered} results from Ookla\textsuperscript{\tiny\textregistered}~\cite{ookla} that we use as a proxy. 

We characterize NPIs adherence using a publicly available dataset from Meta's Data for Good program that provides high spatio-temporal resolution information about mobility changes~\cite{rangemaps}.
Aggregated mobility indicators, obtained from digital crumbs we leave while using or simply carrying mobile phones in our pockets, together with ad-hoc surveys, have been used to characterize adherence with NPIs at the population level~\cite{perra2020non}. The scale of the emergency and the strictness of the measures disrupted such a broad range of activities and behaviors that a simple comparison of aggregated mobility volumes (obtained from mobile phones) with respect to a pre-pandemic baseline shows marked differences and allows to quantify adoption of many types of NPIs. 

We focus our analyses at the municipal level on three Latin American countries: Colombia, Ecuador, and El Salvador. Latin America is often regarded as the most unequal region in the world~\cite{lac_inequality, Lancet2020}. The profound disparities that afflict the region are also reflected in high rates of infection and deaths observed during the Pandemic~\cite{Lancet2021} as well as in the access to and the use of digital tools~\cite{internet_users_lac, wbinternetsubs}. Furthermore, analyses conducted within these countries show large spatial heterogeneities and urban-rural divides in many indicators including digital literacy, skills, and access to broadband~\cite{covid19_digitalinfra}. Unfortunately, these points, together with the limited number of studies in this region of the world, make Latin American countries good case studies to investigate and expand our knowledge about NPIs adoption. 

We find that, during 2020, NPIs significantly affected mobility in the three countries, causing a maximum drop in movements, from pre-pandemic baselines, of $53\%$ in Colombia and $64\%$ in Ecuador and El Salvador. Through a correlation analysis, we first show that municipalities with access to better connectivity also feature more consistent mobility reductions. Such association is preserved when controlling for possible confounders. We estimate that, for every $10$ Megabits per second (Mbps) increase in average fixed download speed, movement reduction increases by another $13\%$, $4\%$, $19\%$ in, respectively, Colombia, Ecuador, and El Salvador. The intensity of the correlation, furthermore, is modulated by the stringency of restrictions imposed. We then adopt a regression approach that, besides digital infrastructure quality, accounts also for many other factors possibly influencing mobility reductions in each municipality. We find that digital infrastructure quality is still a significant predictor of NPIs adoption. Population size, density, and socio-economic status are also associated with higher mobility reductions.

Notwithstanding clear progress made, much is still unknown about the adherence to NPIs especially when it comes to the effects of different socio-economic barriers, such as internet access and quality. We aimed to extend the literature on the subject offering a quantitative investigation about the role of digital infrastructure, along with many other variables, in NPIs compliance during the first wave of COVID-19 Pandemic in three Latin American countries. Understanding such phenomenon is key to informing policies aimed at increasing the resilience to external shocks as well as the equality in communities, cities, and countries.

\section{Results}

\subsection{Proxies of NPIs adoption}

\begin{figure}
    \centering
    \includegraphics[width=\textwidth]{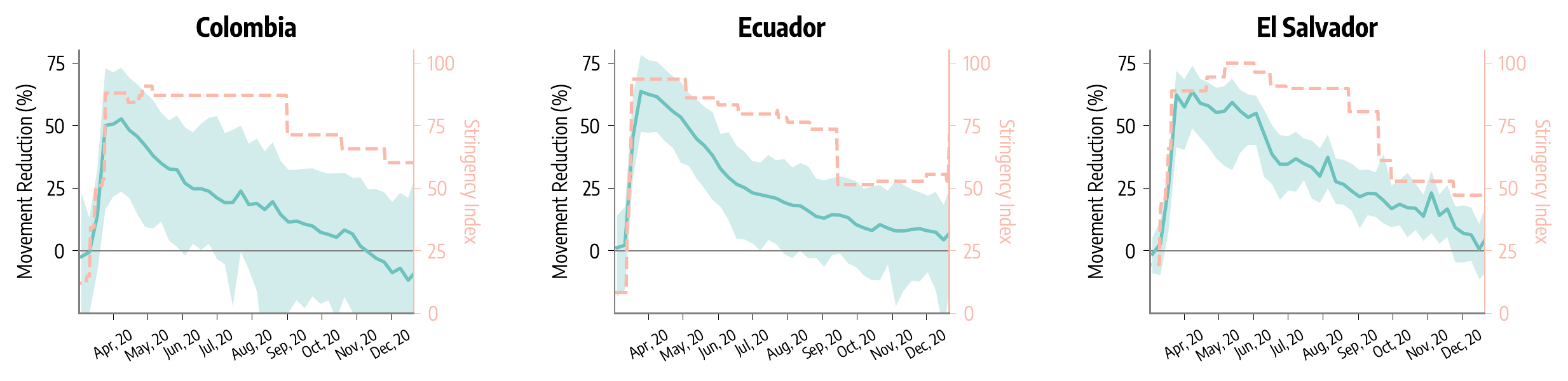}
    \caption{\textbf{Mobility reductions following the establishment of NPIs in Colombia, Ecuador, and El Salvador.} We show the \textit{movement reduction} metric between 2020/03/01 and 2020/12/31 for Colombia, Ecuador, and El Salvador. We show national average (solid line) and the minimum-maximum interval (shaded area) computed over all municipalities in the three countries for which we have data. We also show the stringency index (orange dashed line) of policies implemented to curb COVID-19 spread in the three countries. Horizontal line indicates the pre-pandemic mobility baseline}
    \label{fig:range_maps}
\end{figure}

We characterize NPIs adoption by quantifying how aggregated mobility changed during 2020 in the three countries. To this end, we use the \textit{Movement Range Maps} from Meta's Data for Good program~\cite{rangemaps}. This dataset is publicly available and provides the percentage reduction in movement observed with respect to a pre-pandemic baseline (\textit{movement reduction}). Data are available for different municipalities and have temporal resolution of the day. In Fig.~\ref{fig:range_maps} we show the weekly evolution of the mobility reduction throughout 2020 for Colombia, Ecuador, and El Salvador. We compute the national average (solid line) and the minimum-maximum interval (shaded area) across all municipalities for which we have data. Together, we also report the Stringency Index from the Oxford COVID-19 Government Response Tracker~\cite{Hale2021} which measures the strictness of policies implemented to contain COVID-19 in the three countries (see Sec.~\ref{sec:data}). During March 2020 mobility dropped sharply reaching a maximum reduction in late March/early April ($53\%$ for Colombia, $64\%$ for Ecuador and El Salvador). This drop matches the introduction of increasingly tougher measures, as shown by the evolution of the Stringency Index in the three countries. After reaching the maximum, we observe an inversion in the mobility which anticipates, and then follows, the relaxing of NPIs in early summer. The movement reduction approaches the pre-pandemic baseline (i.e., $\sim 0\%$ reduction) by late 2020.  

It is important to note how heterogeneities within each country do exist. At the peak of mobility drops in Colombia, for example, the municipality associated with the minimum change shows a mobility reduction of $24\%$. In the same week, the municipality featuring the strongest decline has a reduction of $73\%$, marking an absolute minimum-maximum difference of $49\%$. We note similar, though less pronounced, patterns also for Ecuador and El Salvador (minimum-maximum absolute difference of, respectively, $31\%$ and $25\%$). 

The \textit{Movement Range Maps} dataset provides also a second mobility metric describing the fraction of individuals that appear to stay within a small area for the whole day. The two mobility metrics provide different pieces of information, but they can be considered complementary. The overall change in mobility is generally used in similar research~\cite{perra2020non}. It provides a more comprehensive and nuanced view about the impact on NPIs on mobility allowing, for example, to appreciate tendencies towards shorter and less frequent trips. Therefore, in the main text, we will carry out analyses only for the \textit{movement change} metric, while we refer the reader to the Supplementary Information for the analogous analyses for the second mobility metric. The overall picture emerging from both metric is coherent.

\subsection{Proxies of digital infrastructure quality}

We measure the quality of the digital infrastructure using as a proxy internet Speedtest\textsuperscript{\tiny\textregistered} results provided by Ookla\textsuperscript{\tiny\textregistered}~\cite{ookla}. Ookla's Speedtest Intelligence\textsuperscript{\tiny\textregistered} solution offers analysis of internet access performance metrics, such as connection data rate~\cite{feamster2020measuring}. The tests are geolocalized and provide download/upload speed (expressed in Megabits per second) and latency (in milliseconds) for fixed networks (see more details in Sec.~\ref{sec:data}). For the purpose of this analysis, we characterize the quality of digital infrastructure using fixed download speed. Our dataset includes more than $90$M Speedtest\textsuperscript{\tiny\textregistered} measurements among the three countries. 

\begin{figure}
    \centering
    \includegraphics[width=\textwidth]{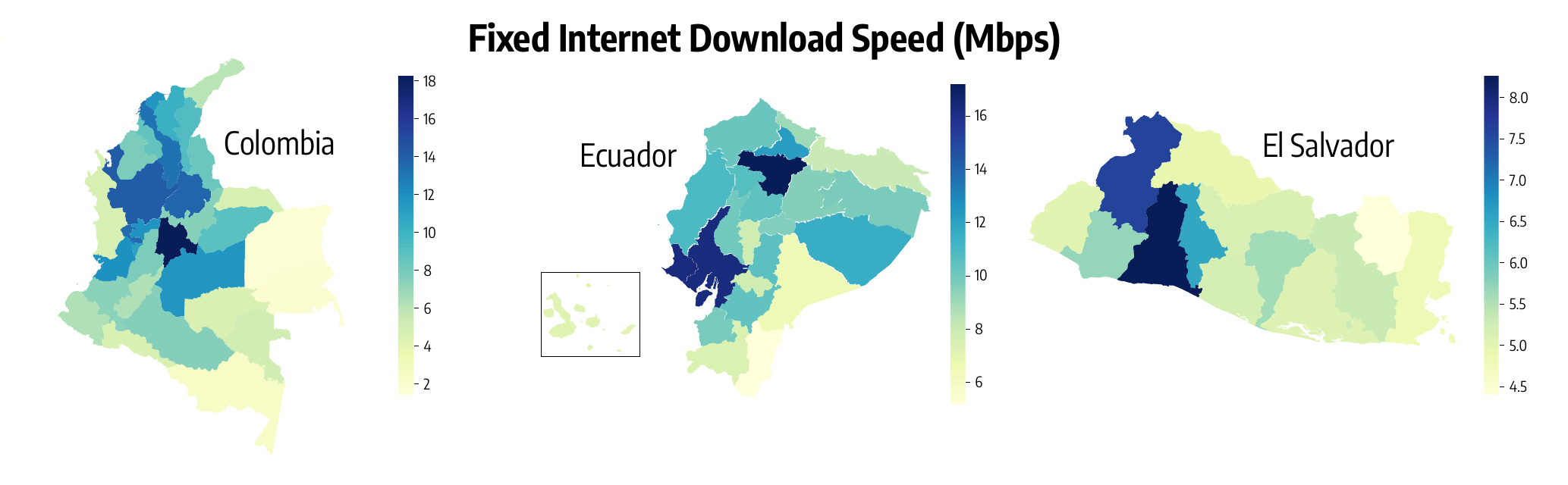}
    \caption{\textbf{Fixed internet quality in Colombia, Ecuador, and El Salvador (2019/2020).} Average fixed internet download speed (expressed in Mbps) computed from Ookla\textsuperscript{\tiny\textregistered} Speedtest Intelligence\textsuperscript{\tiny\textregistered} data for different departments of Colombia, Ecuador, and El Salvador.}
    \label{fig:ookla_maps}
\end{figure}

In Fig.~\ref{fig:ookla_maps} we show the average download fixed internet speed (expressed in Megabits per second) in different departments of Colombia, Ecuador, and El Salvador. From the plot we observe that the quality of digital infrastructure varies widely across regions within the same country. In El Salvador, the average download speed in \textit{La Libertad} is nearly double that in \textit{Morazán}. In Colombia, the department with the best infrastructure shows a $\sim 13$ times higher download speed with respect to the one with the worst network. In Ecuador, the ratio between higher and lower speed among departments is $\sim 3$.

Considering the emphasis of such metric in our analysis, it is fundamental to notice some limitations of testing speed as a proxy of the digital infrastructure. Indeed, the outcome of an individual's test might differ from the real internet speed. A recent review on the subject highlights some of the key issues~\cite{feamster2020measuring}. Some are rather technical aspects linked to bottlenecks in home networks (i.e., Wi-Fi routers), the details of the software or tool used to make a measurement, and the number of devices sharing the connection. Others factors instead are linked to the users themselves. In fact, their tests might not reflect the average speed since they might be done when a user is experiencing connectivity issues or needs to know the level of connectivity available in a area that is new to them. Furthermore, only users more digitally aware know about these types of services. These factors induce undoubtedly some self-selection biases in the sample.
However, Ookla\textsuperscript{\tiny\textregistered} is considered as a \emph{canonical} network performance testing service widely used to infer the features of internet connectivity across and within regions by academic and official institutions ~\cite{vakataki2021visualizing,ford2021form,feamster2020measuring}. Also, our analysis aggregates Speedtest\textsuperscript{\tiny\textregistered} measurements at the level of municipality. In doing so, some of the more technical issues mentioned above are averaged across many different instances. Furthermore, our analyses are not dependent on the absolute value of aggregated connectivity but on the differences across municipalities. Overall, it is important to acknowledge how the data used here is only a proxy of the digital infrastructure.  

\subsection{Association between mobility change and digital infrastructure quality}

Considering the novelty of the metric in this context, before moving to a more rigorous analysis, we first investigate the association between changes in mobility and the quality of the digital infrastructure in different municipalities. 

\begin{figure}
    \centering
    \includegraphics[width=\textwidth]{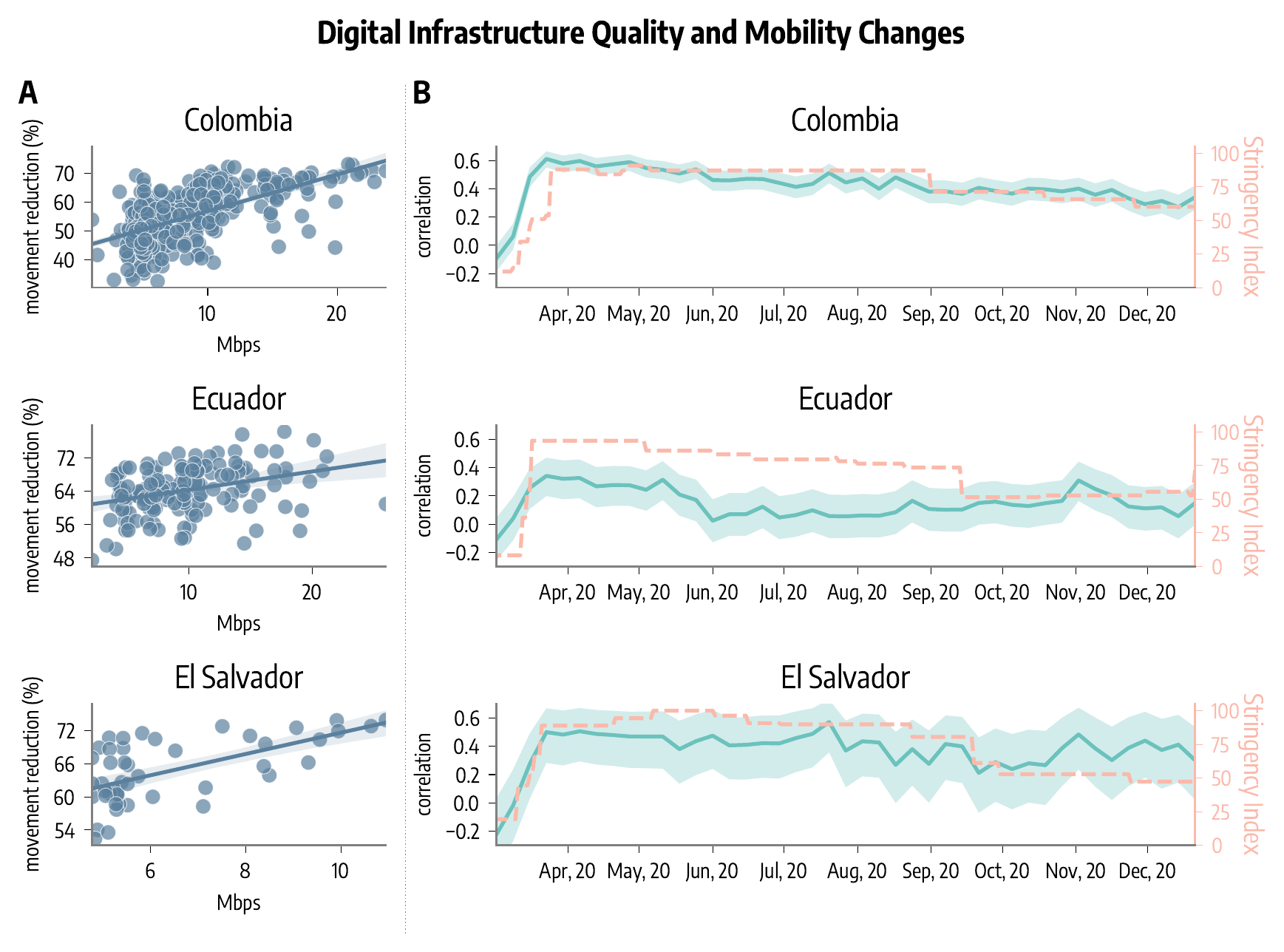}
    \caption{\textbf{Association between mobility reduction and digital infrastructure quality in different municipalities of Colombia, Ecuador, and El Salvador.} A) We plot the greatest reduction in mobility against the average download speed in different municipalities. B) We plot the Pearson correlation coefficient (median and $95\%$ CI) between average weekly movement reduction and average download speed of different municipalities. The orange dashed lines in all plots represent the Stringency Index.}
    \label{fig:correlations_intime}
\end{figure}

In Fig.~\ref{fig:correlations_intime}-A we plot the maximum percentage reduction in mobility (i.e., the greatest level of compliance to NPIs) against the average fixed download speed of each municipality in the three countries. The plot reveals a significant positive linear correlation between the two quantities. Indeed, we obtain Pearson's correlation coefficient of $0.62$ ($95\%$ CI: $[0.55; 0.68]$) for Colombia, $0.34$ ($95\%$ CI: $[0.19; 0.47]$) for Ecuador, and $0.61$ ($95\%$ CI: $[0.39; 0.76]$) for El Salvador. This indicates that $38\%$, $12\%$, and $40\%$  of the variance of the greatest drawdown in movements among municipalities in respectively Colombia, Ecuador, and El Salvador is explained by the quality of the digital infrastructure expressed as average fixed download speed. The slopes of the regression lines reported in the figure indicate that as download speed increases by 10 Megabits, the maximum mobility reduction increases by another $13\%$ in Colombia, $4\%$ in Ecuador, and $19\%$ in El Salvador. Hence, this first finding suggests that municipalities with higher digital infrastructure quality are also those where mobility has changed (reduced) more.

In Fig.~\ref{fig:correlations_intime}-B we show the temporal evolution of the Pearson correlation coefficient between weekly mobility change and average fixed download speed of different municipalities in Colombia, Ecuador, and El Salvador. As expected from the previous analysis, we notice positive correlations. More interestingly, we notice that correlations tend to follow the Stringency Index, shown in the figure as an orange dashed line. Indeed, during early March 2020, when the first restrictions were imposed, correlations have a significant jump, reaching a peak close to the peak of restrictions. Afterward, correlations tend to decrease together with the Stringency Index as some restrictions are partially lifted.

When interpreting this result is important to notice two facts. Correlation does not imply causation. To highlight this aspect we have opted for the use of the word ``association" between the two quantities. Furthermore, there may be an association between these two quantities because both are correlated to other confounders. For example, previous works have shown that the adherence to NPIs of different communities correlates with their socio-economic status~\cite{pullano2020population, Duenas2021, Gozzi2021santiago, fraiberger2020uncovering}. At the same time, better network coverage might be correlated with higher socio-economic conditions. Indeed, internet access and investments in infrastructure favor human and economic growth~\cite{MORARIVERA2021102076, MEDEIROS2021105118, galperin2017internet_poverty}. As a result, movement changes and network quality may show significant correlations because both are directly influenced by the socio-economic status of the municipality. We tackle the issue of confounders in two ways. First, we make use of partial correlations. We compute the Pearson partial correlation between movement reduction and network quality controlling for the socio-economic status of different municipalities~\cite{Kim2015}. We consider the Relative Wealth Index publicly available as part of Meta's Data for Good program as a proxy for the socio-economic status of different municipalities (see Sec.~\ref{sec:data} for details). The correlations presented in Fig.~\ref{fig:correlations_intime}-A, after controlling for the socio-economic status, become $0.32$ ($95\%$ CI: $[0.23, 0.42]$), $0.29$ ($95\%$ CI: $[0.14, 0.43]$), $0.40$ ($95\%$ CI: $[0.13, 0.62]$) for Colombia, Ecuador, and El Salvador respectively. We notice how the partial correlation is smaller but still significant and showing positive sign. We obtain wider confidence intervals in the case of El Salvador due to the smaller sample size. Indeed, we have data for $459$ municipalities in Colombia, $164$ in Ecuador, and $56$ in El Salvador. In the Supplementary Information, we report more details about the methodology of partial correlations and we repeat the analysis using also other features as controls. We find that the overall picture remains unaltered.

As a second and more systematic step, in the next section, we investigate the link between the two quantities by using regression techniques with many additional features to identify the effects and importance of each factor.

\subsection{Regression analysis}

We perform regression analyses to study the extent to which different municipalities changed mobility as a function of several independent variables. 
We first use a static approach in which the dependent variable is the maximum movement reduction observed in 2020 for different municipalities (see Fig.~\ref{fig:correlations_intime}-A). Besides the average download speed ($Mbps$) that we introduced previously and that acts as a proxy of the digital infrastructure quality, we consider several additional independent variables that we can classify in three categories:
\begin{itemize}
    \item socio-demographic features: total population ($population$), population density ($density$), fraction of over $60$ ($60+$);
    \item socio-economic features: GDP per capita ($GDP$), relative wealth index ($RWI$);
 \item epidemiological indicators: total number of cases per 100k reported in the week of the maximum movement reduction ($cases$).
\end{itemize}

\begin{figure}
    \centering
    \includegraphics[width=\textwidth]{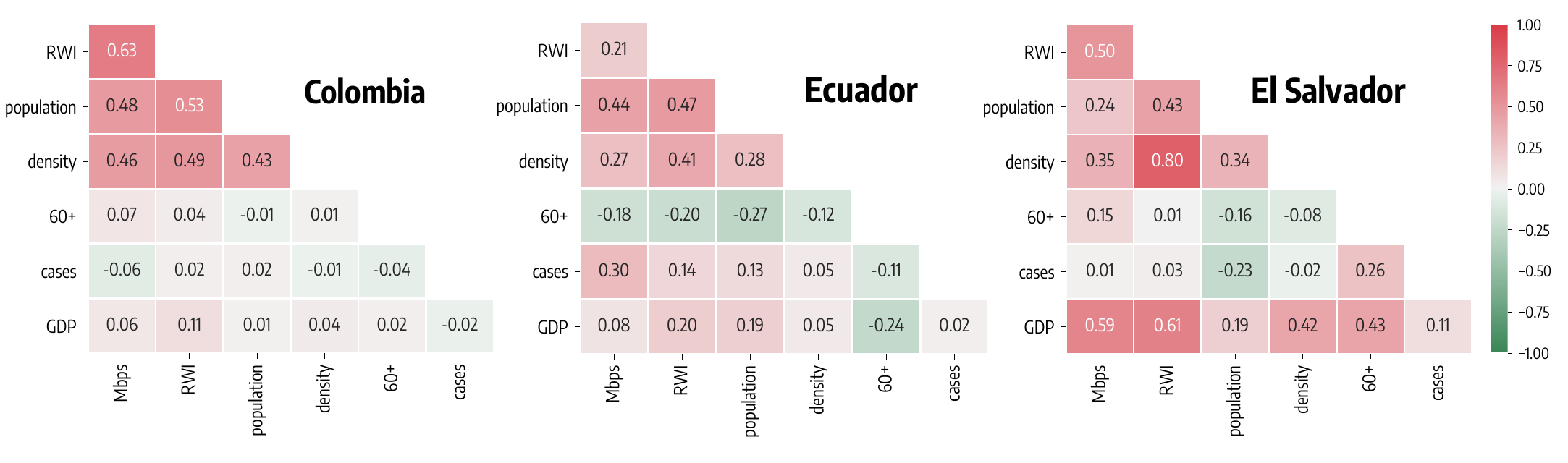}
    \caption{\textbf{Correlations between independent features}.}
    \label{fig:multicoll}
\end{figure}

Before moving forward, we investigate the correlation between independent variables. In Fig.~\ref{fig:multicoll} we show the Pearson correlation between covariates for the three countries. We obtain a maximum absolute coefficient of $0.63$ between $RWI$ and $Mbps$ in Colombia, of $0.48$ between $population$ and $RWI$ in Ecuador, and of $0.80$ between $density$ and $RWI$ in El Salvador. As expected, some of the variables are indeed correlated. In order to access how such correlations might affect the regression we use the maximum variance inflation factor which quantifies the multicollinearity between variables~\cite{Hair.2009}. We obtain $2.01$, $1.59$, and $4.31$ for Colombia, Ecuador and El Salvador which indicate moderately low multicollinearity. Values above $5$ are customarily used to flag multicollinearity problems.

In the main text, we use ordinary least squares, but in the Supplementary Information we repeat analyses using regularized regressions (i.e., ridge regression) and bootstrapping to show the robustness of our results to both the model and the estimation techniques. Regularized regressions in particular are better suited to limit the issues of multicollinearity. The results across all methods used are consistent. All features (independent and dependent) are standardized to their distribution. Full details on data, pre-processing, model and estimation are provided in the Materials and Methods section and the Supplementary Information. 

\begin{figure}
    \centering
    \includegraphics[width=\textwidth]{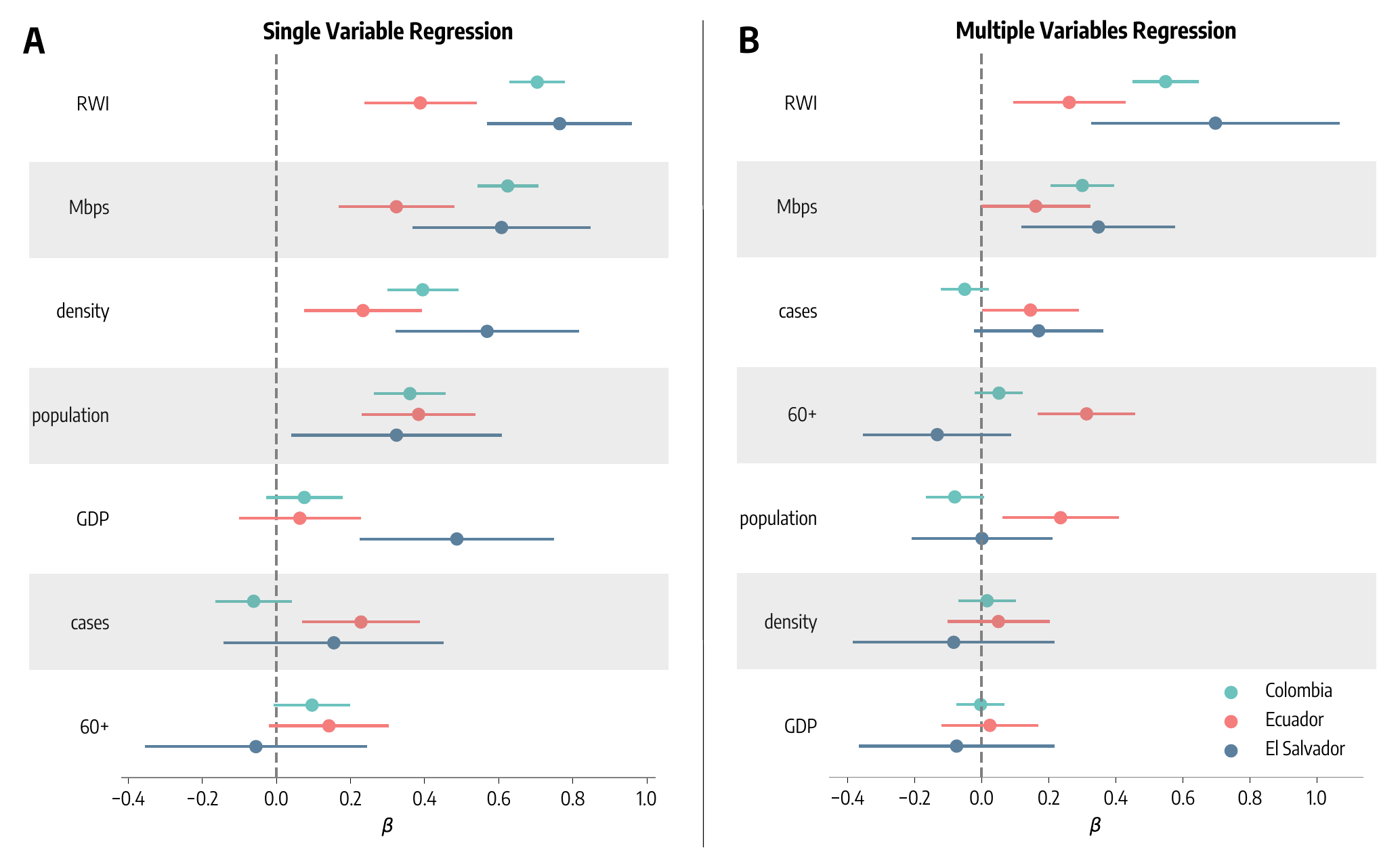}
    \caption{\textbf{Coefficients of the regressions (maximum movement reduction as dependent variable)}. A) Coefficients of the single variable regressions. B) Coefficients of the multiple variables regression.}
    \label{fig:static_coeffs}
\end{figure}

We use both single and multiple variables regressions. In the single variable analysis, we simply regress each feature singularly against the maximum movement reduction of different municipalities. Estimated standardized coefficients (medians and $95\%$ confidence intervals) are reported in Fig.~\ref{fig:static_coeffs}-A. The relative wealth index has a significant positive coefficient in all three countries. The GDP per capita coefficient is also positive but is significant only for El Salvador. This finding confirms the critical role of socio-economic attributes in NPIs adherence reported in previous works~\cite{doi:10.1098/rsif.2020.1035, pullano2020population, Gozzi2021santiago, LEE2021102563, Chang2021_socioecon}. Looking at socio-demographic features, we find that both population and density have a positive significant coefficient in all countries. This is in line with previous findings that have shown how individuals living in larger and denser areas had more pronounced mobility reductions~\cite{doi:10.1098/rsif.2020.1035, gauvin_socioecon}. An older population has been observed to be associated with higher mobility reductions~\cite{doi:10.1098/rsif.2020.1035, gauvin_socioecon}. However, in our results the role of the fraction of $60+$ people is less clear, showing a positive but barely significant coefficient in the case of Colombia and Ecuador. The number of reported cases per 100k has a positive and significant effect only for Ecuador. This is line with previous research where higher attack rates were found to be associated with stronger mobility reductions~\cite{gauvin_socioecon}. The small influence of reported cases may be because the highest mobility drop was reached in different municipalities during March/April 2020 following the introduction of restrictions and grim news coming from other countries rather than local COVID-19 incidence. As a result, most of the municipalities reported very few or no cases in the week when mobility reduction reached its acme. Finally, the single variable regression analysis confirms the significant positive role of average download speed previously underlined through the correlation analysis.

In the multiple variables analysis, the maximum movement reduction is regressed against all features previously introduced (coefficients are reported in Fig.~\ref{fig:static_coeffs}-B). The strong importance of the socio-economic status is confirmed by the positive significant coefficient of the relative wealth index in all three countries. Across the board, the $RWI$ shows the highest standardized coefficient, therefore it represents the most important feature among those considered. Interestingly, the coefficient of the average download speed remains positive and significant in all three countries. This result, combined with the partial correlations, confirms the importance of digital infrastructure in NPIs adherence even after controlling for several other factors. Other attributes have in general smaller importance when considered in concert with the others. The number of cases has a marginally significant and positive effect only for Ecuador and El Salvador. Population size and fraction of $60+$ remain significant only in the case of Ecuador. Overall the regression model has a coefficient of determination ($R^2$) of $0.56$, $0.34$, and $0.69$ in respectively Colombia, Ecuador, and El Salvador. 

As an additional assessment on the role of average download speed, we build a second model by repeating the multiple variables regression without this variable. Then, we compute the Akaike Information Criterion (AIC) of the two models as $AIC = 2 k - 2 log(L)$, where $k$ is the number of independent variables and $L$ is the likelihood function~\cite{akaike1998information}. Therefore, the $AIC$ considers both model's complexity and the accordance between model's estimates and observations. The difference of the AICs of the model with and without download speed as independent variable ($\Delta_{AIC}$) is $-38.4$, $-2.23$, $-8.5$ for Colombia, Ecuador, and El Salvador. As a rule of thumb, a decrease in the AIC of at least $2$ units indicates that the model with lower $AIC$ is significantly better~\cite{Portet2020}. Turning the $\Delta_{AIC}$ into the relative likelihood as $exp(\Delta_{AIC}/2)$ we get that the model without download speed is, for the three countries, $<10^{-3}$, $0.32$, and $0.01$ times as probable as the model with download speed. This additional finding confirms the influential role of the quality of digital infrastructure in NPIs compliance.

\begin{figure}
    \centering
    \includegraphics[width=\textwidth]{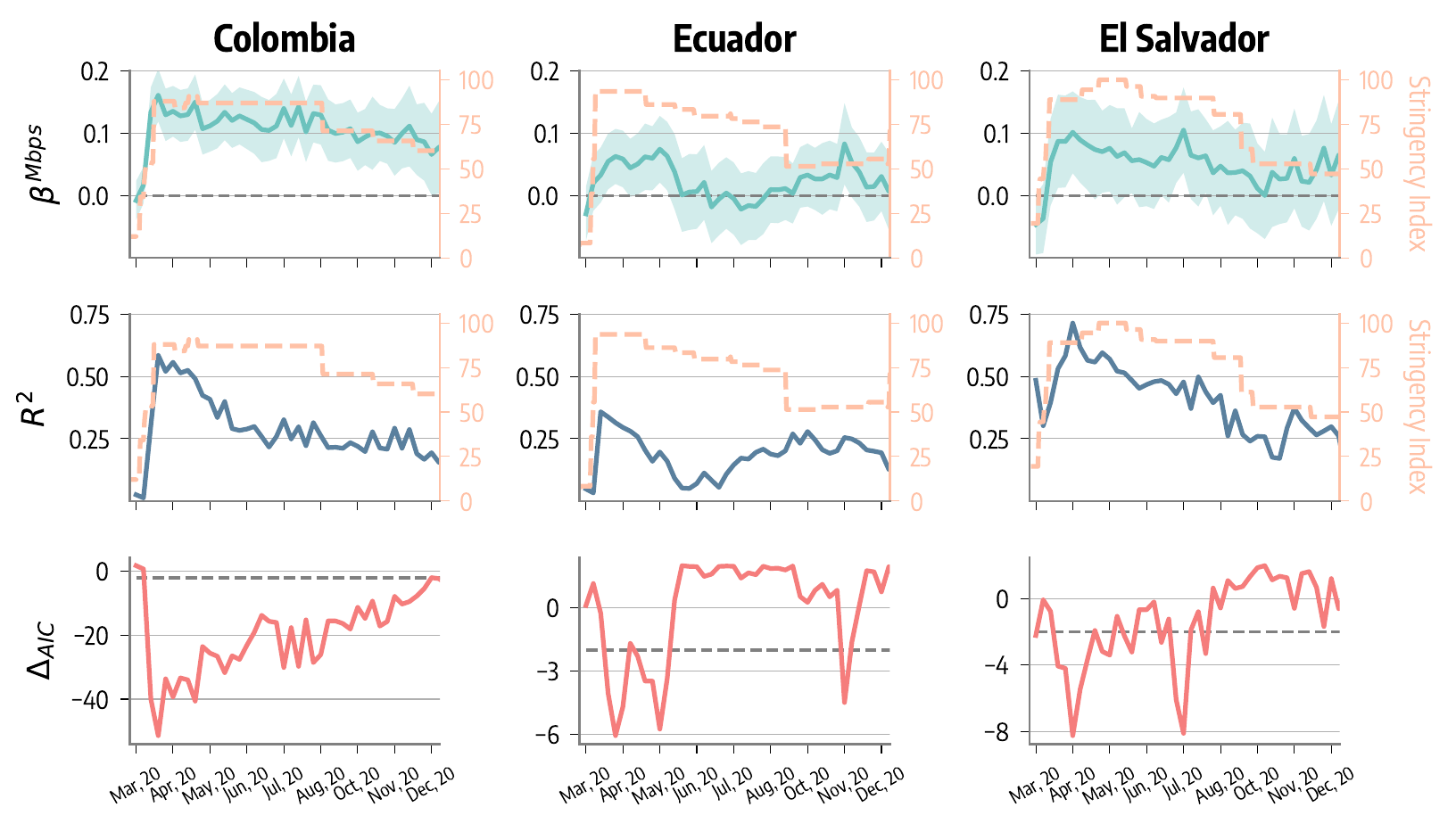}
    \caption{\textbf{Weekly regression results}. First line shows the evolution of the regression coefficient of the average download speed as a function of time for the three countries. Second line shows the coefficient of determination $R^2$ in different weeks. Third line shows the evolution of the $\Delta_{AIC}$ between the model with and without download speed as an independent variable. We average all quantities over a month to rule out the influence of noise. The orange dashed line in the different figures represents the evolution of the Stringency Index.}
    \label{fig:time_coeffs}
\end{figure}

In addition to the static approach, we propose a second regression that explores possible time-varying relationships between dependent and independent variables. For each week of 2020, we repeat the multiple variables regression using however as dependent variable the average reduction in mobility observed in that week across the different municipalities. Also, the independent variable $cases$ in this cases refers to the number of weekly cases reported per $100k$ in that week. In Fig.~\ref{fig:time_coeffs} we show the evolution in time of the estimated coefficient for the average download speed ($\beta^{Mbps}$). We observe a similar trend across the three countries. Indeed, $\beta^{Mbps}$ is $\sim 0$ at the beginning of March, 2020, then it grows and reaches a maximum around late March/April, concurrently with the peak of restrictive measures (Stringency Index reported in figure as an orange dashed line). In the second row of Fig.~\ref{fig:time_coeffs}, we find again a common pattern across the three countries: the coefficient of determination $R^2$ tends to follow the stringency of NPIs measured with the Stringency Index. This indicates that the model is better at predicting mobility changes when measures implemented are stricter. Finally, the third row of the figure reports the time evolution of the $\Delta_{AIC}$ between the model with and without $Mbps$ as independent variable. The figure confirms the importance of the variable, especially during the early months of 2020, where we obtain $\Delta_{AIC}$ values that are smaller than $-2$.

\section{Conclusions}

The burden of the COVID-19 Pandemic and of the measures put in place to fight it are distributed unequally across different social, economic, and demographic backgrounds~\cite{perra2020non, pullano2020population, Duenas2021, Gozzi2021santiago, fraiberger2020uncovering, Menaeabg5298, Chang2021, topriceanu2020inequality,  munoz2020racial, yi2020health, mathur2020ethnic}. In this work, we studied the relationship between NPIs compliance and its determinants during 2020 in the municipalities of Colombia, Ecuador, and El Salvador. Besides previously studied factors - such as socio-economic status and epidemiological indicators - we explored the role of a barrier affecting NPIs adherence: the quality of the digital infrastructure. Analyzing a unique dataset including tens of millions of geolocalized internet Speedtest\textsuperscript{\tiny\textregistered} metrics we showed that municipalities with a better connection were associated with higher mobility reductions. We found that this correlation remains significant also after controlling for confounders, including the socio-economic level of the municipality. By using several regression models we explored the role of other independent variables. We found that mobility reductions were more pronounced in larger, denser, and wealthier municipalities. In addition, these analyses confirmed that the digital infrastructure quality explains part of the differences in NPIs compliance and that its effects are higher at the peak of the strictness of non-pharmaceutical interventions. 

These findings corroborate and expand the literature. Previous studies explored - mainly using surveys - the role of internet access in the engagement in specific activities such as telemedicine and homeschooling during the Pandemic. They showed, for example, that poor Internet connectivity at home explains gaps in student performances~\cite{Bauer2020OvercomingMH} and that lack of access to digital tools poses a barrier to the adoption of digital health innovations and contributes to poor health outcomes~\cite{info:doi/10.2196/19361}. Interestingly, these works found that these effects remain significant also after controlling for socio-economic attributes. A report focused on the early months of the Pandemic in the US, showed results that are compatible with our findings~\cite{NBERw26982}. The authors used a binary variable to describe the access to high-speed internet and showed that this is a relevant factor for compliance to NPIs following state-level mandates.

The present work comes with limitations. Heterogeneity in the usage of mobile devices can lead to biases in aggregated mobility data such as the Movement Range Maps~\cite{doi:10.1098/rsif.2012.0986, 10.1371/journal.pcbi.1003716, 10.1145/3442188.3445881, schlosser2021biases}. The socio-demographic and socio-economic features we used are only aggregated and do not have access to individual information. 
As explained before, also Ookla\textsuperscript{\tiny\textregistered} data presents limitations. Indeed, the result of an individual's test might differ from the real internet speed~\cite{feamster2020measuring}. However, here we do not aim to provide an accurate representation of the quality of the digital infrastructure but rather characterize the differences between municipalities. Therefore, thanks to the relative nature of the analyses, key issues of tests do not represent a concern. Another possible bias may be introduced by the heterogeneous testing behavior of individuals. Indeed, some may perform a Speedtest\textsuperscript{\tiny\textregistered} much more frequently than others. We solve this by averaging results from a single user before further aggregation, as explained in the Materials and Methods section. While we controlled for several factors known to play a role in NPIs adherence, part of the results may be also explained by omitted variables. On top of these, we mention information about formal employment and labor structure~\cite{doi:10.1098/rsif.2020.1035} that, because of lack of data, we did not model explicitly. In the Supplementary Information, we explore this dimension by repeating analyses with additional features for Colombia. We find that even after including labor formality and relative importance of different economic sectors among the independent variables of the regressions, digital infrastructure quality is still a significant factor explaining NPIs adherence. We only considered the quality of digital infrastructure. Arguably, also the adoption of digital tools is an important factor playing a role in NPIs adherence. Because of the scarce availability of updated data on digital tools adoption (especially for El Salvador and Ecuador) at the spatial level of our analysis (i.e., municipalities), in this work, we decided to focus only on the quality of the digital infrastructure and its possible impact on NPIs adherence. Nonetheless, the two aspects are not unrelated. In the case of Colombian municipalities, for example, we find a strong positive correlation between infrastructure quality and internet penetration. In the Supplementary Information, we follow this direction studying the role of digital tools adoption in NPIs compliance proposing as proxies the number of Speedtest\textsuperscript{\tiny\textregistered} measurements performed (for all the three countries) and the number of internet subscriptions (only for Colombia) in different municipalities. We find that, together with digital infrastructure quality, also the adoption of digital tools is a significant determinant of NPIs adherence. Finally,  we are aware of spatial heterogeneities, at the levels of municipalities and regions, in NPIs that might have affected communities within each country differently ~\cite{doi:10.1098/rsif.2020.1035}. However, we do not have detailed data reconstructing the timeline of NPIs at this resolution. For this reason, while we expect the stringency of NPIs to affect the mobility within each municipality, we did not include such variables in the regression. We leave this extension to future work.

In conclusion, our work confirms the role of socio-economic and socio-demographic factors in NPIs compliance and sheds light on possible barriers to adoption represented by the heterogeneous digital infrastructure quality and the digital divide. Our results call for policies and targeted investments aimed at closing the digital gap, improving network reliability as well as equality across communities.

\section{Materials and Methods}

\subsection{Datasets}
\label{sec:data}

\noindent \textbf{Oxford COVID-19 Government Response Tracker.} We consider the Stringency Index from the Oxford COVID-19 Government Response Tracker to quantify the strictness of policies implemented to curb SARS-CoV-2 spread~\cite{Hale2021}. The index goes from a minimum of $0$ (i.e., no measure) to a maximum of $100$ (i.e., strictest measures). It is computed as the average of several sub-indicators that describe, for example, the level of closure of schools and workplaces, restrictions to international travel, stay at home requirements, and the implementation of public health information campaign. For more details on the methodology refer to Ref.~\cite{oxford_codebook}. The Stringency Index has daily resolution and is available at the country level for Colombia, Ecuador, and El Salvador. \\

\noindent \textbf{Movement Range Maps.} To estimate adherence to NPIs we consider changes in the mobility captured in the dataset \textit{Movement Range Maps} from Meta's Data for Good Program~\cite{rangemaps}. This dataset is publicly available and has been used to characterize changes in mobility in several works in the context of COVID-19 Pandemic~\cite{Menaeabg5298, Kishore2021, Cortes2021}. It provides two main metrics: i) a percentage reduction in movement computed with respect to a pre-pandemic baseline, and ii) the fraction of individuals that appear to stay within a small area for the whole day. The two metrics are computed using de-identified data of Facebook users who opt-in for Location History and background location collection (see Ref.~\cite{rangemaps_methods} for more details on the methodology). The data are available for several countries at the municipal level (GADM2) and have a temporal resolution of the day. We have data for $459$ of the $1065$ GADM2 areas (i.e., municipalities) in Colombia, for $164$ of the $223$ in Ecuador, and for $56$ of the $266$ in El Salvador. The spatial distribution of municipalities for which data are available is shown in the Supplementary Information. We have data for at least one municipality in each of the $32$, $24$, and $14$ GADM1 areas (i.e., regions) in, respectively, Colombia, Ecuador, and El Salvador. In the analyses presented above, we performed different aggregations of this data. In particular, weekly statistics were computed excluding weekends, and only municipalities for which the whole week Monday to Friday was available were included. National statistics were simply computed over all the municipalities of the country for which data were available. \\

\noindent \textbf{Relative Wealth Index.} We characterize the socio-economic status of different GADM2 areas using the Relative Wealth Index (RWI) from Meta's Data for Good Program. The index, publicly released in 2021, measures the relative standard of living within countries using machine learning, de-identified connectivity data, satellite imagery, and other nontraditional data sources (more details in Ref.~\cite{chi2021microestimates}). It is available for nearly $93$ low and middle income worldwide at a very high spatial resolution ($30m$ population density tiles). Here, we aggregate the RWI at the municipal level taking the weighted average according to the population of tiles within each municipality. A full description of this aggregating procedure is provided by in Ref.~\cite{rwi_calculation}. In the Supplementary Information, we compare the RWI against more traditional measures of wealth for the countries considered. We find that the RWI is highly correlated to those measures. \\

\noindent \textbf{Speedtest\textsuperscript{\tiny\textregistered} data.} We characterize the quality of digital infrastructure using as proxy Speedtest Intelligence data\textsuperscript{\tiny\textregistered} by Ookla\textsuperscript{\tiny\textregistered}~\cite{ookla}. Speedtest\textsuperscript{\tiny\textregistered} apps offer free analyses of Internet access performance metrics, such as connection data rate. The tests are geolocalized and provide download/upload speed (expressed in Megabits per second), and latency (in milliseconds) for fixed network. The dataset includes $65'863'831$ tests for Colombia, $4'821'878$ for El Salvador, and $22'159'403$ for Ecuador performed during 2019-2020. Before computing any statistics we clean the data as follows. First, we exclude all tests showing $0$ \textit{Mbps} download speed since these are generally tests that failed and are not informative of the real network quality. Second, to remove the influence of outliers we exclude tests that show a download speed higher than $2$ Gigabits per second, as this is considered as the maximum value for broadband technology. Lastly, we exclude tests that show a latency higher than the $95^{th}$ percentile. The cleaning process leads to the exclusion of around $5\%$ of the total tests. After preprocessing, we take the median of the results of tests performed by a single user. This is done to avoid over-representation of users using the service more often than others. Finally, for each municipality, we compute the median download speed of all tests performed in that area. \\

\noindent \textbf{Population data.} For Colombia we get the population of different municipalities from the official census~\cite{popDANE}. Because of the lack of updated data on municipality population in Ecuador and El Salvador, we consider the publicly available \textit{High-Resolution Population Density Maps and Demographic Estimates} from Meta's Data for Good program~\cite{density_maps}. This dataset also provides the number of people over $60$ for all three countries. \\

\noindent \textbf{COVID-19 cases.} We consider official epidemiological sources for Colombia~\cite{colombia_covid19cases}, Ecuador~\cite{covid_cases_ecuador}, and El Salvador~\cite{elsalvador_covid19cases}. \\

\noindent \textbf{GDP data.} We get GDP data from Ref.~\cite{dane_gdp} for Colombia, from Ref.~\cite{hdi_ecuador} for Ecuador, and from Ref.~\cite{almanaque262} for El Salvador. \\

\noindent \textbf{Geographic data.} We download spatial data for the three countries and their sub-divisions from the Database of Global Administrative Areas~\cite{gadm}. \\

\section*{Author's contributions}
All authors designed the study, wrote and approved the manuscript. N.G. performed the analyses.

\section*{Funding}
The research is partially funded by The World Bank Digital Economy for Latin America and Caribbeans (DE4LAC) initiative. The findings, interpretations, and conclusions expressed in this paper are entirely those of the authors. They do not necessarily represent the views of the International Bank for Reconstruction and Development/World Bank and its affiliated organizations, or those of the Executive Directors of the World
Bank or the governments they represent.

\section*{Acknowledgements}

All authors thank the High Performance Computing facilities at Greenwich University, Ookla\textsuperscript{\tiny\textregistered}, The World Bank and the Development Data Partnership. All authors thank James Carroll, Natalija Gelvanovska-Garcia, Mykhailo Koltsov, Katherine Macdonald, Carolina Mejia-Mantilla for their support and review. N.G. acknowledges support from the Doctoral Training Alliance. N.G. thanks Thomas Rabensteiner and Zsófia Zádor for the useful discussions.

\section{Supplementary Information}

\subsection{Relative Wealth Index}
\begin{figure}[ht!]
    \centering
    \includegraphics[width=\textwidth]{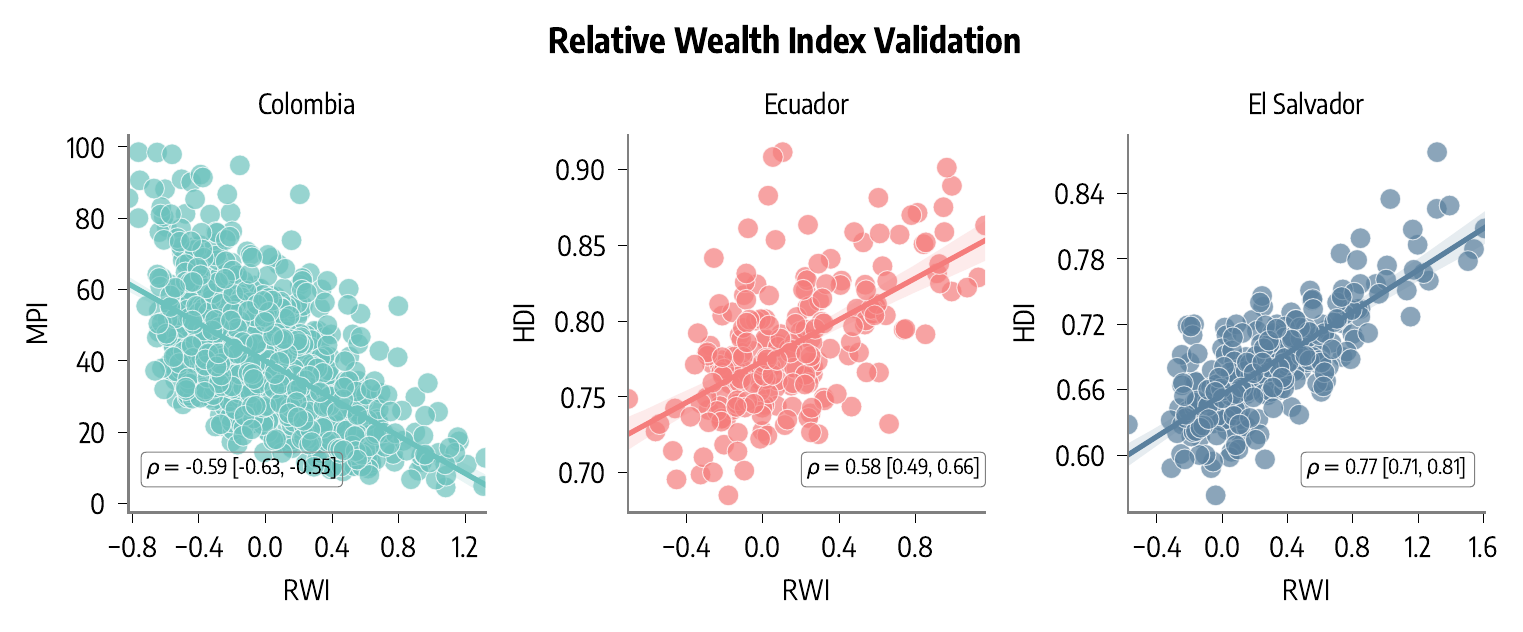}
    \caption{\textbf{Validation of Relative Wealth Index.} We show the correlation of the relative wealth index with the multidimensional poverty index (Colombia) and the human development index (Ecuador and El Salvador).}
    \label{fig:rwi_validation}
\end{figure}

In Fig.~\ref{fig:rwi_validation} we provide a comparison of the Relative Wealth Index ($RWI$) against more traditional measures of wealth. More in detail, we compare it to the multidimensional poverty index ($MPI$) for Colombia (2018)~\cite{dane_mpi} and to the Human Development Index ($HDI$) of municipalities in Ecuador (2016)~\cite{hdi_ecuador} and El Salvador (2011)~\cite{almanaque262}. Although these measures have different definitions and express different pieces of information, in both cases we find a high significant Pearson correlation coefficient between the measures: $\rho=-0.59$ ($[-0.63, -0.55]$) in the case of Colombia, $\rho=0.58$ ($[0.49, 0.66]$) in Ecuador, and $\rho=0.77$ ($[0.71, 0.81]$) in El Salvador. The negative sign obtained for Colombia is expected, indeed the $RWI$ is a measure of wealth while the $MPI$ is a measure of poverty. 

\subsection{Geographic coverage of Range Maps}
\begin{figure}[ht!]
    \centering
    \includegraphics[width=\textwidth]{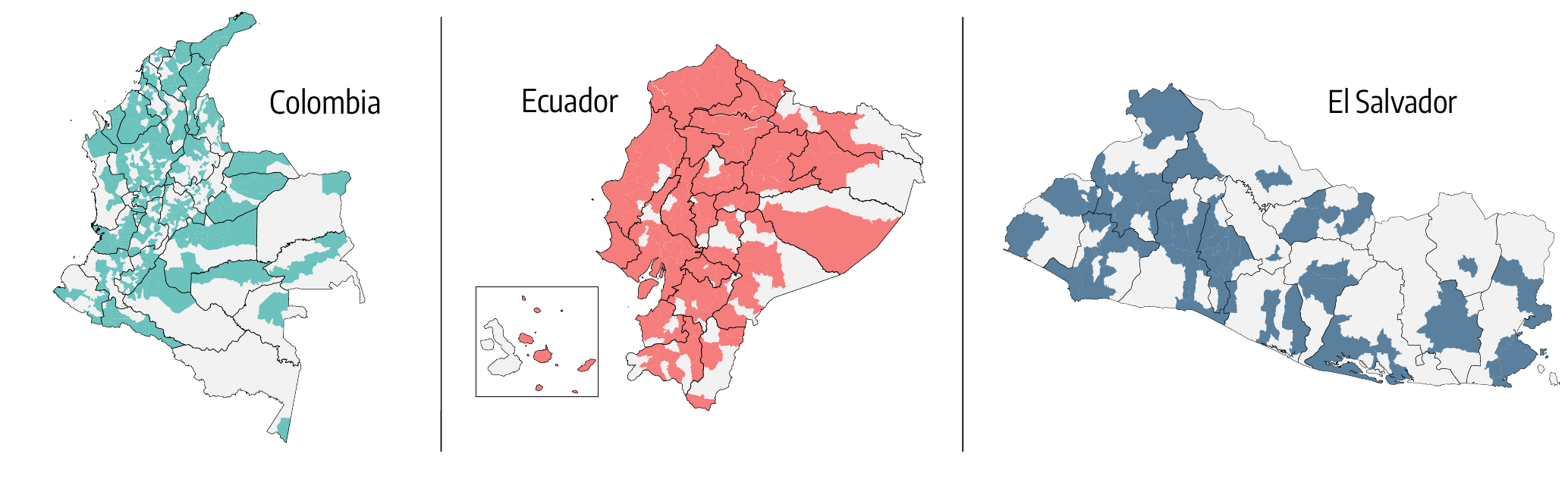}
    \caption{\textbf{Geographic coverage of Range Maps in Colombia, Ecuador, and El Salvador.}}
    \label{fig:rangemaps_coverage}
\end{figure}

In Fig.~\ref{fig:rangemaps_coverage} we show the geographic coverage of the Meta's Range Maps in the three countries of focus, Colombia, Ecuador, and El Salvador. Municipalities for which data are not available are colored in grey. We have data for $459$ of the $1065$ GADM2 areas (i.e., municipalities) in Colombia, for $164$ of the $223$ in Ecuador, and for $56$ of the $266$ in El Salvador. More in detail, at least one municipality in each of the $32$, $24$, and $14$ GADM1 areas (i.e., regions) in the three countries is included in the dataset.

\subsection{Partial correlations}
Consider three random variables $X$, $Y$, and $Z$. Suppose that both $X$ and $Y$ are correlated to $Z$ and that we are interested in finding the association between $X$ and $Y$ accounting for the fact that are both correlated to $Z$. Partial correlation answers to this problem: indeed, it can be regarded as the association between two variables ($X$ and $Y$) after removing the effect of a third variable ($Z$)~\cite{GVK330798693}. In practice, partial correlation is obtained by computing the correlation between the residuals of the two regressions $X$ on $Z$ and $Y$ on $Z$. Like other correlation coefficients, the partial correlation assumes values between $–1$ (perfect negative association) and $+1$ (perfect positive association). In this work, we used the Python library \textit{pingouin} to compute partial correlations~\cite{Vallat2018}.

In the main text, we used partial correlations to find the association between the maximum movement reductions and the average download speed of different municipalities controlling for their socioeconomic status using the Relative Wealth Index. Here we repeat this analysis using several other features as control, namely the GDP per capita, the fraction of $60+$, the population size, and the population density. Results are reported in Tab.~\ref{tab:partial_corr}. The first row shows the simple Pearson correlation coefficient without any control. Each of the following rows reports the partial correlation using different attributes as control. The last row shows the partial correlation using all the previously listed features as control. We see that, also after controlling for several other features, the residual correlation between movement reduction and digital infrastructure quality remains positive and significant in all cases.

\begin{table}[ht!]
\centering
\begin{tabular}{rccc}
\hline
\multicolumn{1}{l}{} & \textbf{Colombia}     & \textbf{Ecuador}      & \textbf{El Salvador}  \\ \hline
\textit{/}           & 0.62 {[}0.55; 0.68{]} & 0.34 {[}0.19; 0.47{]} & 0.61 {[}0.39; 0.76{]} \\
\textit{RWI}         & 0.32 {[}0.23; 0.41{]} & 0.29 {[}0.14; 0.43{]} & 0.4 {[}0.13; 0.62{]}  \\
\textit{GDP}         & 0.62 {[}0.55; 0.68{]} & 0.32 {[}0.17; 0.46{]}  & 0.45 {[}0.19; 0.66{]} \\
\textit{60+}         & 0.59 {[}0.52; 0.65{]} & 0.27 {[}0.12; 0.41{]} & 0.57 {[}0.34; 0.74{]} \\
\textit{density}     & 0.54 {[}0.46; 0.61{]} & 0.29 {[}0.14; 0.43{]} & 0.53 {[}0.29; 0.71{]} \\
\textit{population}  & 0.59 {[}0.52; 0.65{]} & 0.28 {[}0.12; 0.42{]} & 0.58 {[}0.35; 0.74{]} \\
\textit{all}         & 0.31 {[}0.22; 0.4{]}  & 0.25 {[}0.09; 0.39{]} & 0.45 {[}0.18; 0.66{]} \\ \hline
\end{tabular}
\caption{Partial correlations between maximum movement reduction and average download speed in different municipalities using several attributes as control.}
\label{tab:partial_corr}
\end{table}

\subsection{Regularized linear models and bootstrapping}
\subsubsection{Ridge regression}
Ridge regression is a regularized linear model for analyzing data suffering from multicollinearity~\cite{10.2307/1271436}. Indeed, when the independent variables are highly correlated, the standard errors estimated via ordinary least squares may be affected. More in detail, parameters $\beta_j$ are estimated minimizing the following objective function:
\begin{equation}
    \mathcal{L}(\alpha) = \sum_{i=1}^{N} (y_i - \sum_{j=1}^{M}x_{ij} \beta_j)^{2}+ \alpha \sum_{j=1}^{M}\beta_j^2
\label{eq:ridge_loss}
\end{equation}
Where $N$ is the number of samples and $M$ is the number of independent features. With respect to the typical loss function of ordinary least squares, $\mathcal{L}(\alpha)$ has an additional L2 regularization term which is the sum of the squares of the weights multiplied by a parameter $\alpha$. This parameter modulates the importance of the regularization term. When $\alpha=0$ we find the typical expression of ordinary least squares loss function, for higher values instead the L2 regularization becomes more important. In this work, the regularization parameter $\alpha$ is calibrated via leave-one-out cross-validation exploring $30$ log-spaced values between $0.1$ and $10$. We use the implementation of Ridge regression in the Python package \textit{scikit-learn}~\cite{scikit-learn}.

\subsubsection{Bootstrap sampling}
The bootstrap method is a technique for the estimation of regression parameters and confidence intervals. This approach consists in iteratively resampling the dataset with replacement and performing the estimation on the resampled data. More formally, consider the matrix of the independent variables $X \in \mathbf{R}^{N \times M}$ and correspondent array of the dependent variable $Y \in \mathbf{R}^{N}$. At each iteration, we sample with replacement $N$ observations from the dataset, obtaining $X^{bootstrap} \in \mathbf{R}^{N \times M}$ and $Y^{bootstrap} \in \mathbf{R}^{N}$. Then, we use $X^{bootstrap}$ and $Y^{bootstrap}$ to estimate the regression parameters $\hat{\beta}_j$. We repeat this procedure for $T=500$ times. In the end, we use the results of different iterations to compute median and $95\%$ confidence intervals for the parameters. This is a general methodology that can be applied to any model, from ordinary least squares to Ridge regression.

\subsubsection{Robustness of findings to model and estimation method}
\begin{figure}[ht!]
    \centering
    \includegraphics[width=\textwidth]{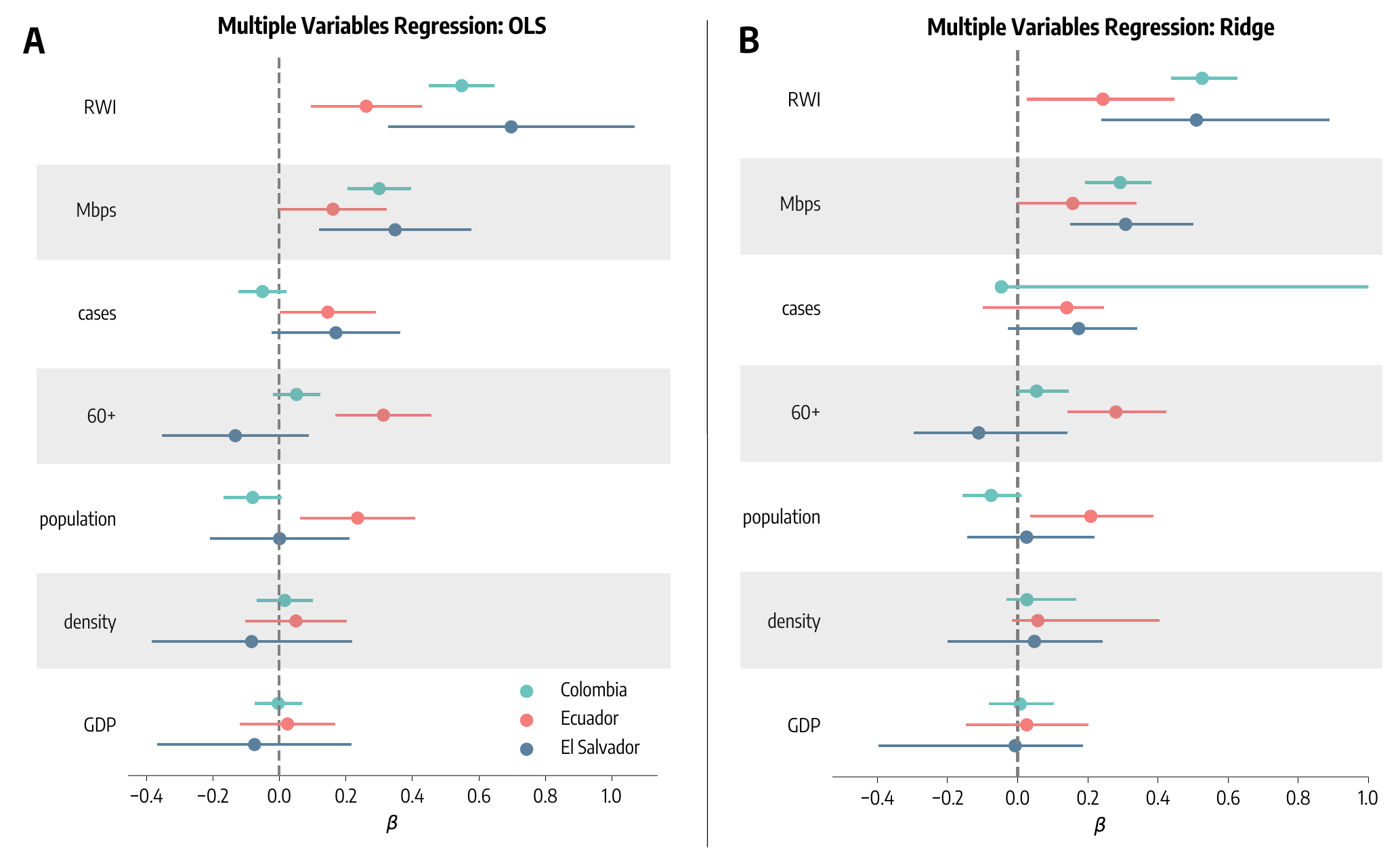}
    \caption{\textbf{Comparison of regression coefficients estimated with ordinary least squares and with Ridge regression and bootstrap sampling.}}
    \label{fig:ridge_ols}
\end{figure}

\begin{figure}[ht!]
    \centering
    \includegraphics[width=\textwidth]{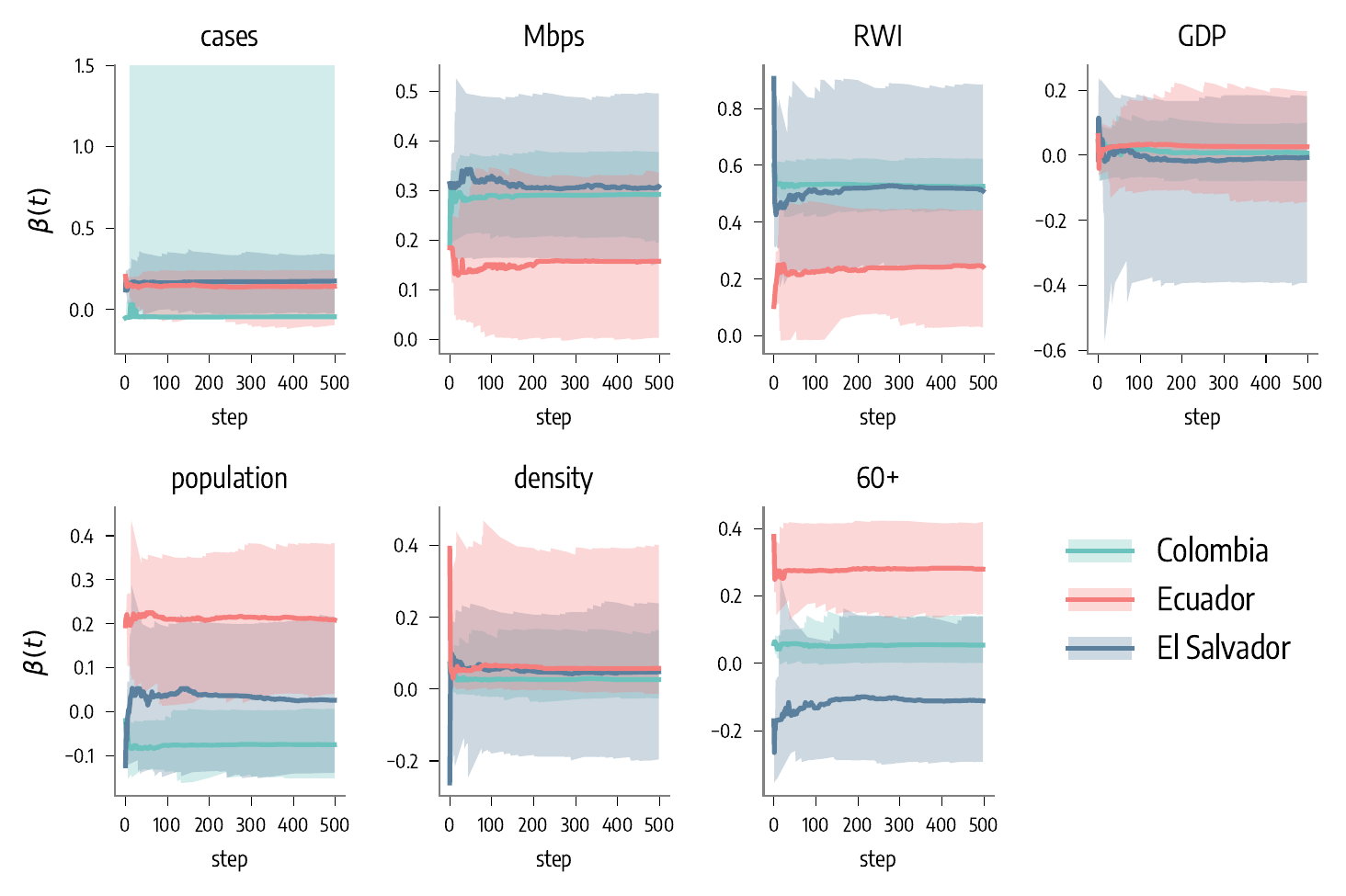}
    \caption{\textbf{Evolution of estimated median and $95\%$ CI as a function of bootstrap step.}}
    \label{fig:bootstrap_conv}
\end{figure}

We repeat the regression presented in the main text in which we used the maximum reduction as the dependent variable and multiple features as independent variables. Instead of ordinary least squares, we use Ridge regression and we perform parameter estimation using the bootstrap method. In Fig.~\ref{fig:ridge_ols} we compare the parameters estimated with OLS (presented in the main text in Fig.5-B) and with Ridge regression. As we can see, parameters obtained with the two approaches are compatible, therefore the overall picture commented in the main text remains valid. We acknowledge wider confidence intervals for the variable $cases$ for Colombia. As mentioned in the main text, the low importance of this variable may be because, at the maximum of mobility reductions, most of the municipalities reported few or no cases, while a small number reported a consistent number of cases. Because of the iterative resampling, the estimation of this parameter may be affected by this distribution. In Fig.~\ref{fig:bootstrap_conv} we show the convergence of parameters estimated via bootstrapping. We plot the estimated medians and $95\%$ as a function of the bootstrap step. We see that $500$ steps are enough for convergence.

\subsection{The \textit{Stay at Home} mobility metric}
\subsubsection{The metric}
The \textit{Movement Range Maps} dataset provides two main metrics, a percentage reduction in mobility with respect to a pre-pandemic baseline (\textit{movement reduction}), and the percentage of users that appear to stay within a small area for the whole day (\textit{stay at home}). In the main text, we characterized the adherence to NPIs with the first metric, here we repeat the analyses using the second one.

\begin{figure}[ht!]
    \centering
    \includegraphics[width=\textwidth]{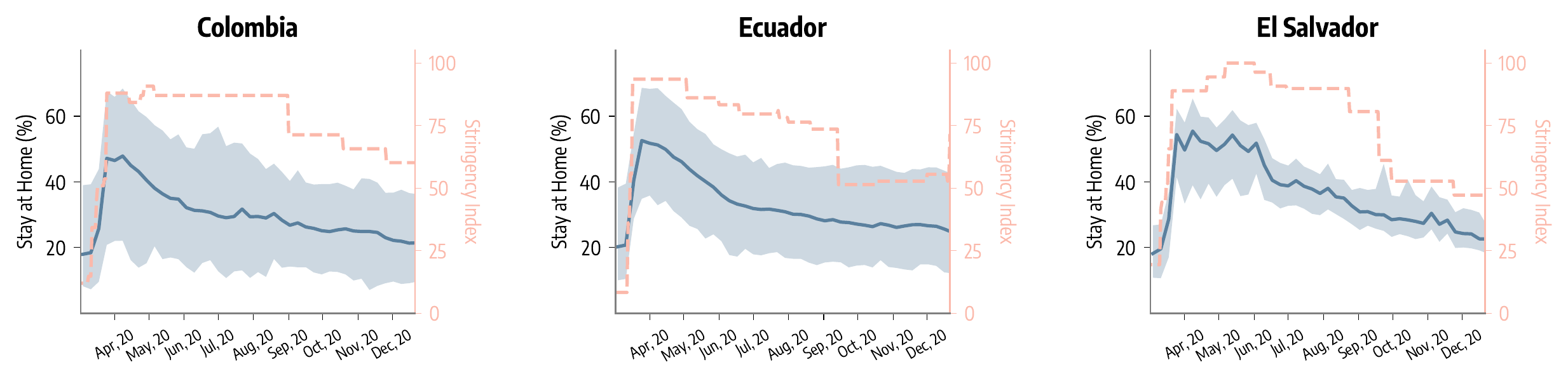}
        \caption{\textbf{Percentage of individuals staying home following the establishment of NPIs in Colombia, Ecuador, and El Salvador.} We show the \textit{stay at home} metric between 2020/03/01 and 2020/12/31 for Colombia, Ecuador, and El Salvador. We show national average (solid line) and the minimum-maximum interval (shaded area) computed over all municipalities in the three countries for which we have data. We also show the stringency index (orange dashed line) of policies implemented to curb COVID-19 spread in the three countries.}
    \label{fig:stayhome}
\end{figure}

In Fig.~\ref{fig:stayhome} we show the weekly evolution of the \textit{stay at home} metric throughout 2020 for Colombia, Ecuador, and El Salvador (national averages as solid lines and minimum-maximum intervals as shaded areas), together with the Stringency Index from the Oxford COVID-19 Government Response Tracker~\cite{Hale2021}. We see that the metric has a similar trend of \textit{movement reduction} presented in the main text. In March 2020 the fraction of people staying at home increased reaching a maximum in late March/early April ($48\%$ for Colombia, $53\%$ for Ecuador, and $55\%$ for El Salvador). After reaching the maximum, we observe an inversion, and the fraction of people remaining at home slowly starts to decrease. Also in this case we find that the mobility metric follows the evolution of the stringency index

\subsubsection{Correlations}
\begin{figure}[ht!]
    \centering
    \includegraphics[width=\textwidth]{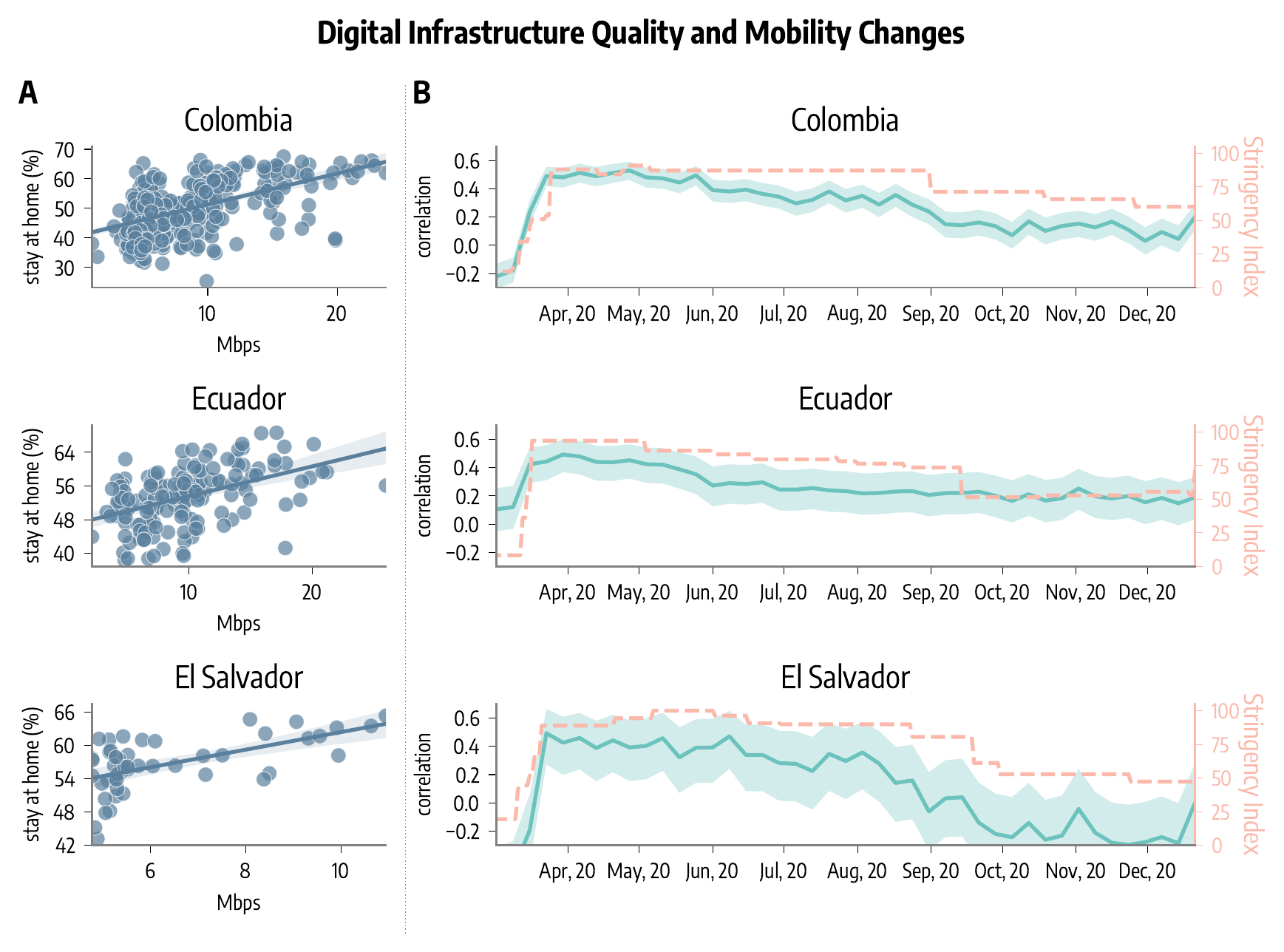}
    \caption{\textbf{Association between fraction staying at home and digital infrastructure quality in different municipalities of Colombia, Ecuador, and El Salvador.} A) We plot the greatest fraction of people staying at home against the average download speed in different municipalities. B) We plot the Pearson correlation coefficient (median and $95\%$ CI) between average weekly \textit{stay at home} metric and average download speed of different municipalities. The orange dashed lines in all plots represent the Stringency Index.}
    \label{fig:corrs_stayhome}
\end{figure}

In Fig.~\ref{fig:corrs_stayhome}-A we plot the average fixed download speed of municipalities against their maximum percentage of people staying at home during 2020. We find for all three countries positive and significant correlations ($0.54$ $[0.47 - 0.61]$, for Colombia, $0.47$ $[0.33 - 0.58]$ for Ecuador, and $0.58$ $[0.36 - 0.74]$ for El Salvador). Furthermore, these coefficients remain significant also after controlling for the socioeconomic status of municipalities. Indeed, we find the following partial correlations using the Relative Wealth Index as a control: $0.27$ $[0.17 -0.36]$, $0.43$ $[0.30 - 0.56]$, and $0.37$ $[0.10 - 0.60]$ for, respectively, Colombia, Ecuador, and El Salvador.

In Fig.~\ref{fig:corrs_stayhome}-B we show the evolution of the Pearson correlation coefficient between the average download speed and the weekly fraction of people staying home during 2020. Also in this case we find that the correlation follows the evolution of the stringency index (orange dashed line in the figure).

\subsubsection{Regressions}
\begin{figure}[ht!]
    \centering
    \includegraphics[width=\textwidth]{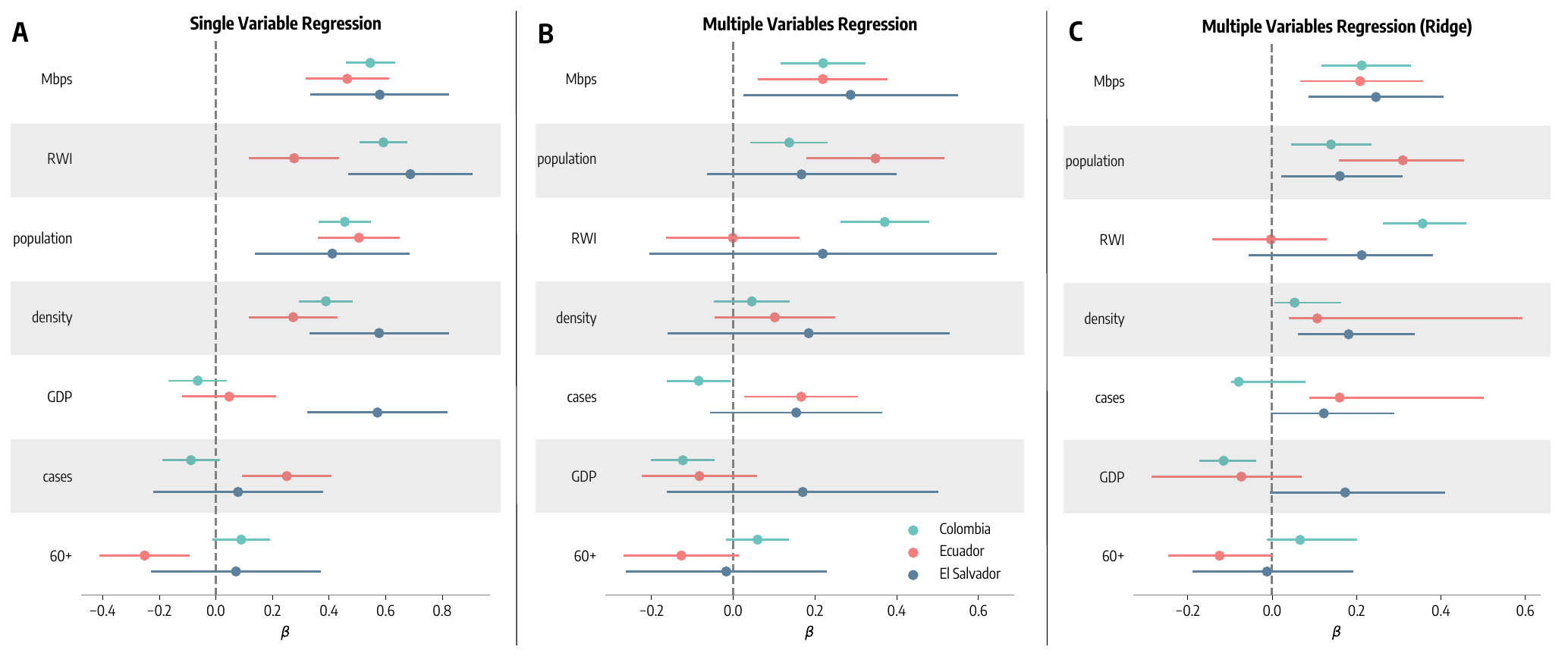}
    \caption{\textbf{Coefficients of the regressions (maximum \textit{stay at home} as dependent variable)}. A) Coefficients of the single variable regression. B) Coefficients of the multiple variables regression. C) Coefficients of the multiple variables regression using Ridge and bootstrapping for estimation.}
    \label{fig:coeff_static_stayhome}
\end{figure}

In Fig.~\ref{fig:coeff_static_stayhome} we show results of the static regression analysis using the maximum fraction of individuals staying at home as the dependent variable. 

Fig.~\ref{fig:coeff_static_stayhome}-A shows the results of the single variable regression. We find very similar results to those of the main text. Higher density and population are associated with stronger adherence to NPIs. Socioeconomic features are also important, indeed the coefficient of the $RWI$ is positive and significant in all three countries while the $GDP$ is significant only for El Salvador. A higher number of reported cases is significantly associated with higher NPIs compliance in Ecuador. Finally, as expected from the correlation analysis, the association between the percentage of people staying home and digital infrastructure quality is positive and significant in all three countries. 
The role of average download speed is confirmed also in Fig.~\ref{fig:coeff_static_stayhome}-B, where the results of the multiple variables regression are reported. Indeed, the coefficient of this variable remains positive and significant. Overall, we get wide confidence intervals. This may be due to multicollinearity between independent features. Indeed, in Fig.~\ref{fig:coeff_static_stayhome}-C we repeat the multiple regression analysis using Ridge regression and bootstrapping for estimation. As expected, we obtain smaller confidence intervals. Population size and density remain significant predictors of NPIs adherence. The number of cases is significant only in the case of Ecuador and marginally significant for El Salvador. Interestingly, the role of socioeconomic figures is less clear. Indeed, we get that the RWI is positive and significant only in the case of Colombia, while the GDP is marginally significant only for El Salvador.

\begin{figure}
    \centering
    \includegraphics[width=\textwidth]{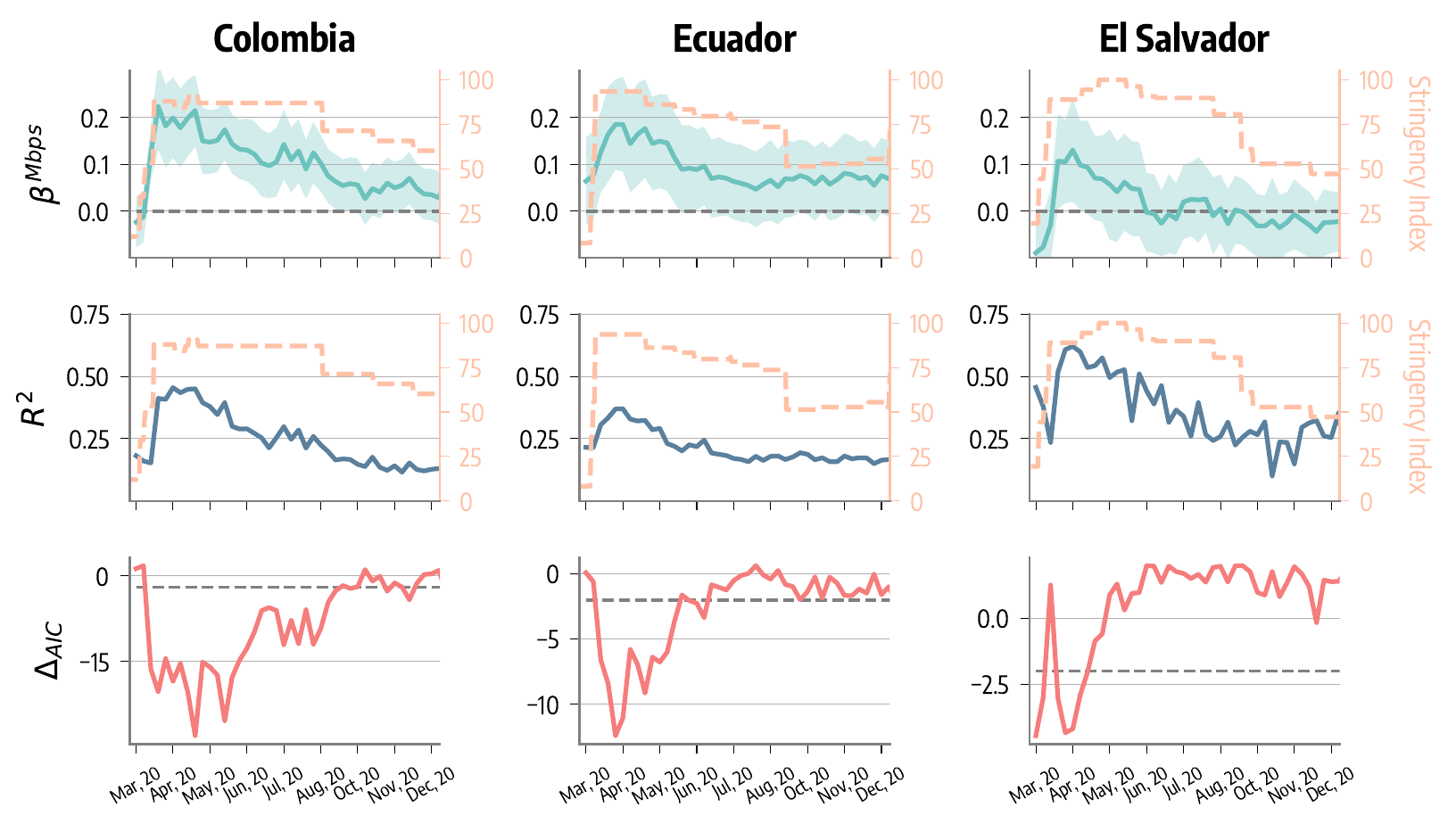}
    \caption{\textbf{Weekly regression results (Stay home as dependent variable)}. First line shows the evolution of the regression coefficient of the average download speed as a function of time for the three countries. Second line shows the coefficient of determination $R^2$ in different weeks. Third line shows the evolution of the $\Delta_{AIC}$ between the model with and without download speed as an independent variable. We average all quantities over a month to rule out the influence of noise. The orange dashed line in the different figures represents the evolution of the Stringency Index.}
    \label{fig:time_coeffs_maxstay}
\end{figure}

Finally, in Fig.~\ref{fig:time_coeffs_maxstay} we show the results of the time-varying regression approach using the \textit{stay at home} as dependent variable. The findings of the main text are confirmed. Indeed, across the three countries the coefficient of internet speed is $\sim 0$ at the beginning of March and then grows reaching a peak in late March/April, concurrently with the maximum strictness of restrictive policies. Also in this case, the coefficient of determination $R^2$ tends to follow the stringency of NPIs measured with the Stringency Index. The third row confirms the importance of the variable $Mbps$, especially during the early months of 2020, where we obtain $\Delta_{AIC}$ values that are smaller than $-2$.

\clearpage
\subsection{Digital tools adoption and NPIs adherence}
\subsubsection{Internet adoption and infrastructure quality}
\begin{figure}[ht!]
    \centering
    \includegraphics[width=\textwidth]{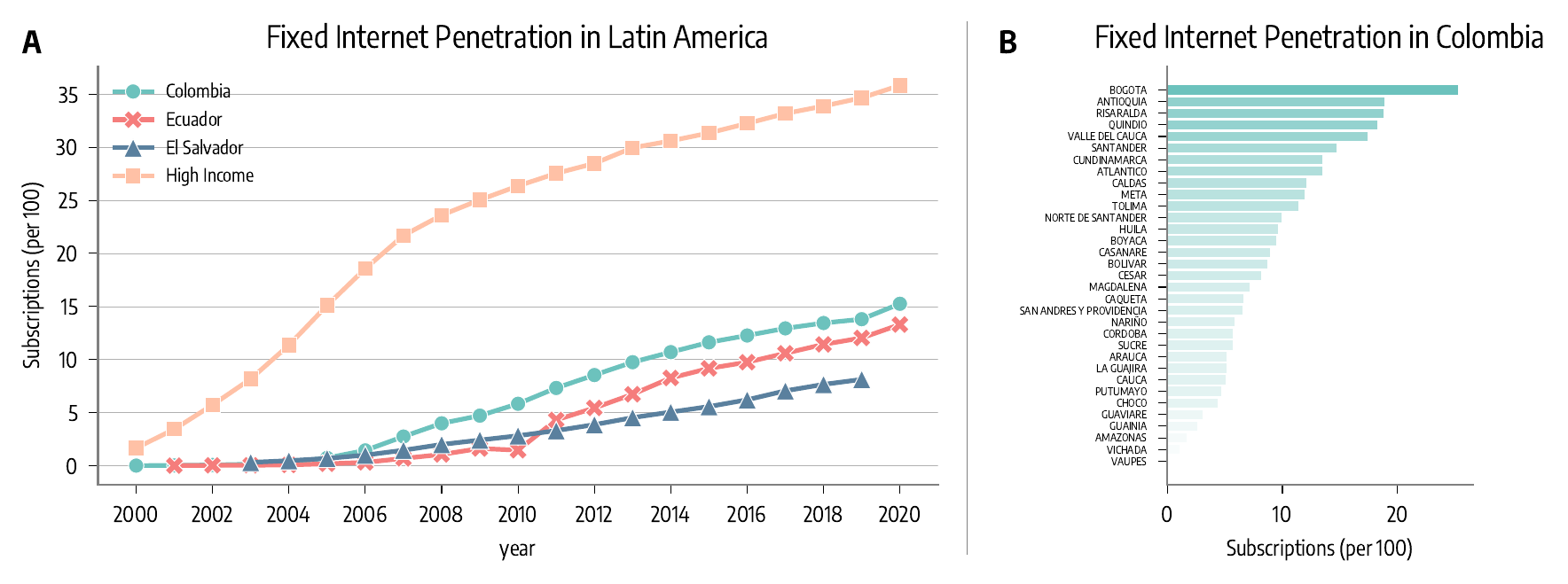}
    \caption{A) Number of fixed internet subscriptions per $100$ in high-income countries, Colombia, Ecuador, and El Salvador. B) Number of fixed internet subscriptions per $100$ in Colombian departments.}
    \label{fig:fixed_subs}
\end{figure}

In the main text, we considered the quality of digital infrastructure as a possible determinant of NPIs adherence. Arguably, also the actual adoption of digital tools is an important factor playing a role in NPIs adherence.
Access to connectivity significantly increased over the past years in Latin America~\cite{internet_users_lac}. This is reflected, for example, in the positive trend observed for the number of fixed internet subscriptions per 100 people~\cite{wbinternetsubs} shown in Fig.~\ref{fig:fixed_subs}-A. Despite this encouraging tendency, the gap with High-Income countries is still worryingly wide. In Colombia and Ecuador, the number of fixed internet subscriptions is comparable to that of other countries in the same region while in El Salvador it is significantly lower. If we push our analysis to a more granular level than general national trends we uncover an additional layer of digital inequality. As an illustrative example, we use data on fixed internet subscriptions per 100 (2019) in the different departments of Colombia~\cite{daneinternet}. The national average is $14.1$ subscriptions per $100$ but this number alone does not communicate the vast heterogeneity across regions. Indeed, as shown in Fig.~\ref{fig:fixed_subs}-B, the internet penetration widely varies across departments, from a maximum of $25.2$ fixed subscriptions per $100$ in the Bogota Department to a minimum of $0.22$ in the Vaupes Department. 
Finally, we mention that internet adoption and quality of connectivity are not unrelated. For example, we find a strong significant correlation between the number of fixed internet subscriptions per $100$ and the related average download speed measured with Ookla\textsuperscript{\tiny\textregistered} Speedtest Intelligence\textsuperscript{\tiny\textregistered} data in the municipalities of Colombia ($\rho=0.63$, $95\%$ CI: $[0.58 - 0.69]$).

\subsubsection{Proxies of digital tools adoption}
As the first proxy for internet adoption, we use the number of fixed internet subscriptions per $100$. To the best of our knowledge, this information is only available at the municipal level for Colombia from Ref.~\cite{daneinternet}. As a second proxy, we consider the number of unique devices per $100$ that performed a Speedtest\textsuperscript{\tiny\textregistered} during 2019-2020. This is more than just a measure of internet adoption. Indeed, among internet users only the more digitally aware know about these types of services. We compute it for the three countries using the Ookla\textsuperscript{\tiny\textregistered} dataset that we used for the calculation of average download speed.

\subsubsection{Correlations between digital tools adoption and NPIs adherence}
\begin{figure}[ht!]
    \centering
    \includegraphics[width=\textwidth]{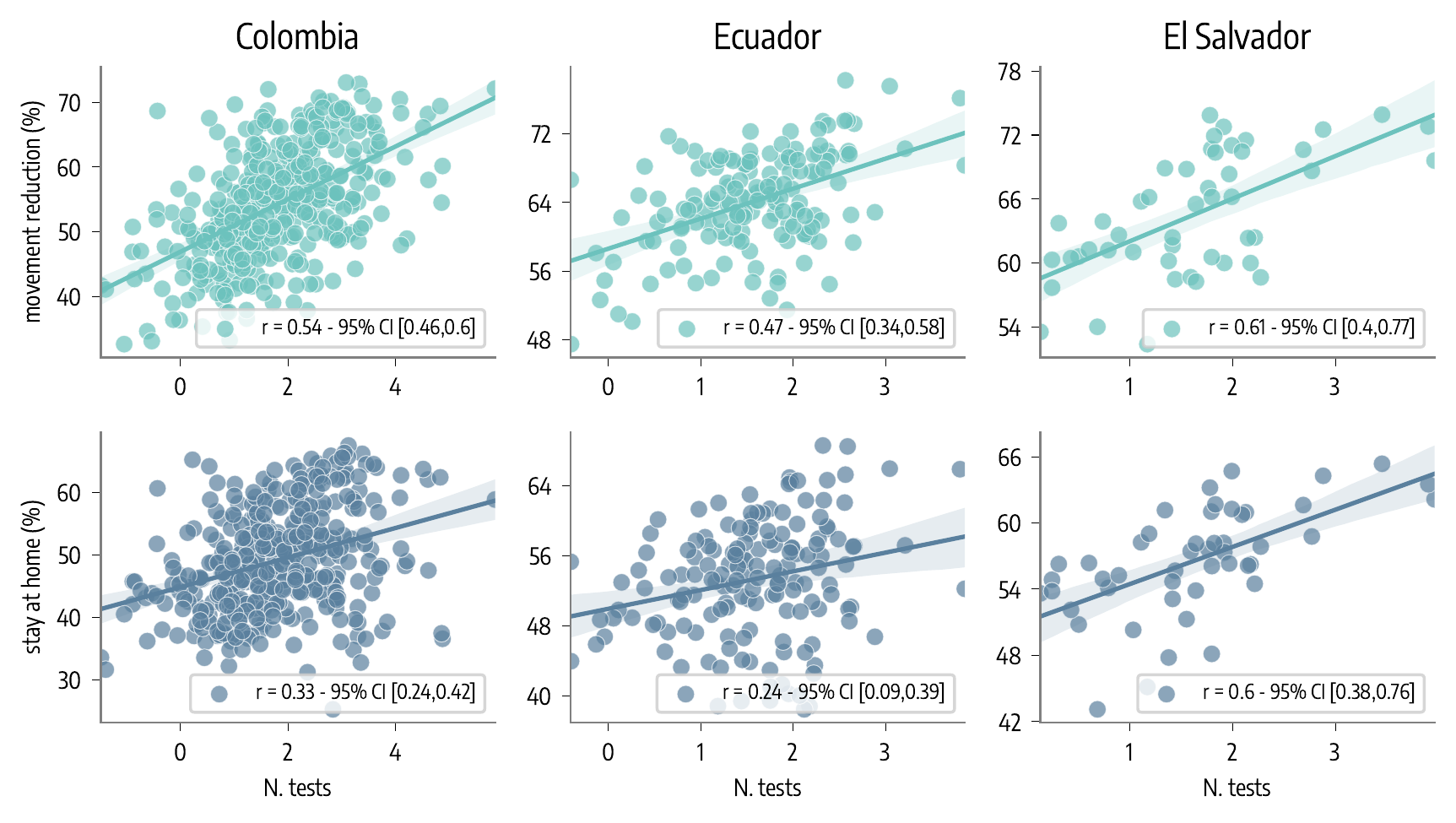}
    \caption{Scatter plot between number of Speedtest\textsuperscript{\tiny\textregistered} measurement and NPIs adherence (described by either the \textit{movement reduction} and the \textit{stay at home} metric).}
    \label{fig:ntest_corrs}
\end{figure}

In Fig.~\ref{fig:ntest_corrs} we plot the number of tests performed per $100$ against NPIs adherence (described by either the \textit{movement reduction} or the \textit{stay at home} metric). To reduce the influence of outliers in the number of tests per capita, we take the logarithm of this quantity. Overall, we find that a higher number of tests is associated with higher compliance. Indeed, the Pearson correlation coefficients reported in the figure vary from a minimum of $0.24$ to a maximum of $0.61$. In Tab.~\ref{tab:partial_corr_ntests} we report the relative partial correlations using the relative wealth index as a control. As expected, the coefficients are smaller, but still significant in most of the cases. Only for El Salvador, we get non-significant or marginally significant coefficients. 

Finally, we repeat the analysis considering the number of fixed internet subscriptions per $100$ rather than the number of tests performed. As explained before, we have this information only for the municipalities of Colombia. We obtain a highly significant correlation coefficient between internet penetration and both maximum movement reduction ($0.65$, [$0.59$; $0.71$]) and the maximum fraction of individuals staying at home ($0.68$ [$0.62$; $0.73$]). The correlations remain significant also after controlling for the socioeconomic status of different municipalities using the relative wealth index (partial correlation of, respectively, $0.29$ [$0.20$; $0.38$] and $0.45$ [$0.37$; $0.52$]).

\begin{table}[ht!]
\centering
\begin{tabular}{rcc}
\hline
\multicolumn{1}{l}{} & \textbf{Movement Reduction} & \textbf{Stay at Home} \\ \hline
\textit{Colombia}    & 0.36 {[}0.27; 0.44{]}       & 0.10 {[}0.0; 0.2{]}   \\
\textit{Ecuador}     & 0.40 {[}0.26; 0.52{]}       & 0.18 {[}0.02; 0.33{]} \\
\textit{El Salvador} & 0.20 {[}-0.09; 0.46{]}      & 0.25 {[}-0.04; 0.5{]} \\ \hline
\end{tabular}
\caption{Partial Pearson correlation coefficients between NPIs adherence and number of tests per 100 using the relative wealth index as control.}
\label{tab:partial_corr_ntests}
\end{table}

\subsubsection{Regression analysis}
We repeat the static regression in which we regress the maximum adherence to NPIs (either maximum \textit{movement reduction} or \textit{stay at home}) against several independent features including the proxies of adoption of digital tools just described. 

\begin{figure}[ht!]
    \centering
    \includegraphics[width=\textwidth]{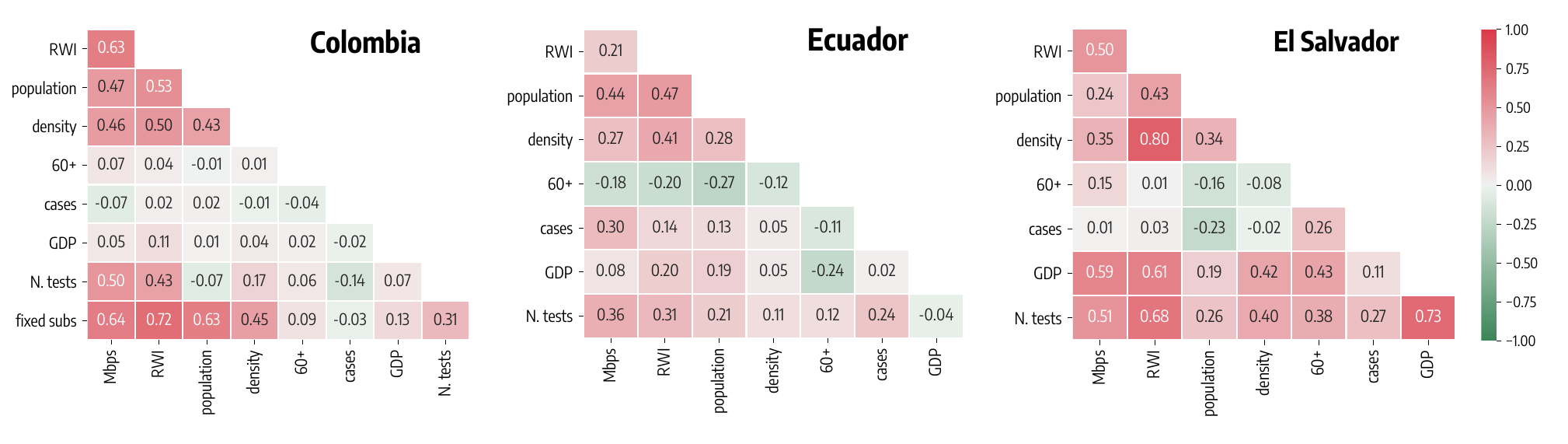}
    \caption{\textbf{Correlations between independent features (proxies of adoption of digital tools included)}}
    \label{fig:multicoll_extrafeatures}
\end{figure}

Before moving to the regression, we estimate the correlations between independent features in Fig.~\ref{fig:multicoll_extrafeatures}. The number of tests performed is generally correlated with the quality of the digital infrastructure ($0.50$, $0.36$, and $0.51$ in the municipalities of Colombia, Ecuador, and El Salvador). At the same time is also correlated with the socioeconomic status of the municipalities. Indeed, the correlation between the number of tests performed and the relative wealth index is, in the three countries, $0.43$, $0.31$, and $0.68$. In El Salvador, it is also correlated with GDP per capita ($0.71$).
The number of fixed internet subscriptions in Colombia is also positively associated with higher download speed ($0.64$) and higher socioeconomic status ($0.72$ with relative wealth index). The maximum variance inflation factors are, respectively, $2.8$, $1.6$, and $5.7$ for Colombia, Ecuador, and El Salvador. Because of these values, we will use Ridge regression and bootstrapping for the estimation of coefficients.

\begin{figure}[ht!]
    \centering
    \includegraphics[width=\textwidth]{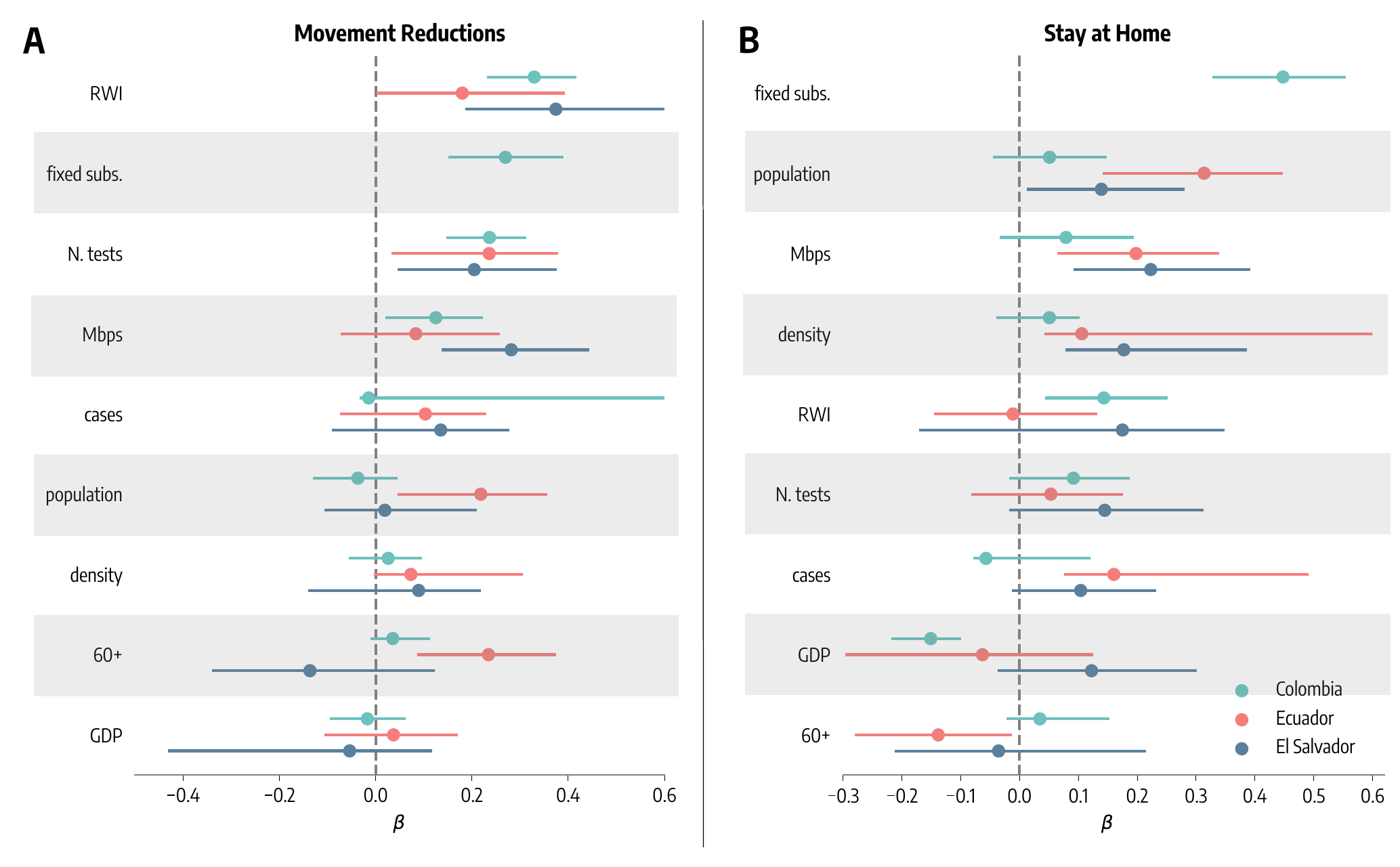}
    \caption{\textbf{Coefficients of the regressions (proxies of adoption of digital tools included).}}
    \label{fig:coeffs_extrafeatures}
\end{figure}

In Fig.~\ref{fig:coeffs_extrafeatures} we report the coefficients of the static regression with the additional features on digital tools adoption included. When the dependent variable is the maximum \textit{movement reduction}, we obtain that the digital infrastructure quality is still a significant predictor for Colombia and El Salvador. In the case of Ecuador, the coefficient is still positive but not significant. We notice that the coefficient of the number of fixed tests performed per $100$ is positive and significant in all three countries. The important role of digital tools adoption is confirmed by the coefficient of the number of fixed internet subscriptions per $100$ in Colombia. 

The picture emerging using the maximum \textit{stay at home} as the dependent variable is consistent. The coefficient of the number of Speedtest\textsuperscript{\tiny\textregistered} measurements is marginally significant for Colombia and El Salvador, while the average download speed remains a significant predictor in the case of Ecuador and El Salvador. We still obtain a highly positive and significant coefficient for the number of fixed internet subscriptions in Colombia. 

Overall, the analysis presented here shows the possible importance of digital tools adoption and confirms the results presented in the main text.

\subsection{Formal employment and labor structure}
We include in our regression data about employment. To the best of our knowledge, this data are available at the municipal level only Colombia, therefore we will restrict our analysis to this country only. 
First, we consider labor formality. This is simply computed dividing the number of individuals formally employed by the number of people in working age ($15-64$) in a given municipality~\cite{o2016path, doi:10.1098/rsif.2020.1035}. Employment data are provided in Ref.~\cite{datlascolombia}, while subnational population statistics in Ref.~\cite{colombia_pop_hdx}. Second, we consider the relative importance of different economic sectors. From Ref.~\cite{dane_gdp} we get the GDP of primary, secondary, and tertiary sector. Then, we compute the share of the total municipal GDP that corresponds to each sector. In the regression we include only the share of GDP associated with primary and tertiary sectors. Indeed, the third variable (i.e., the share of the secondary sector) would be redundant since the sum of the three shares is always $1$.

\begin{figure}[ht!]
    \centering
    \includegraphics[width=\textwidth]{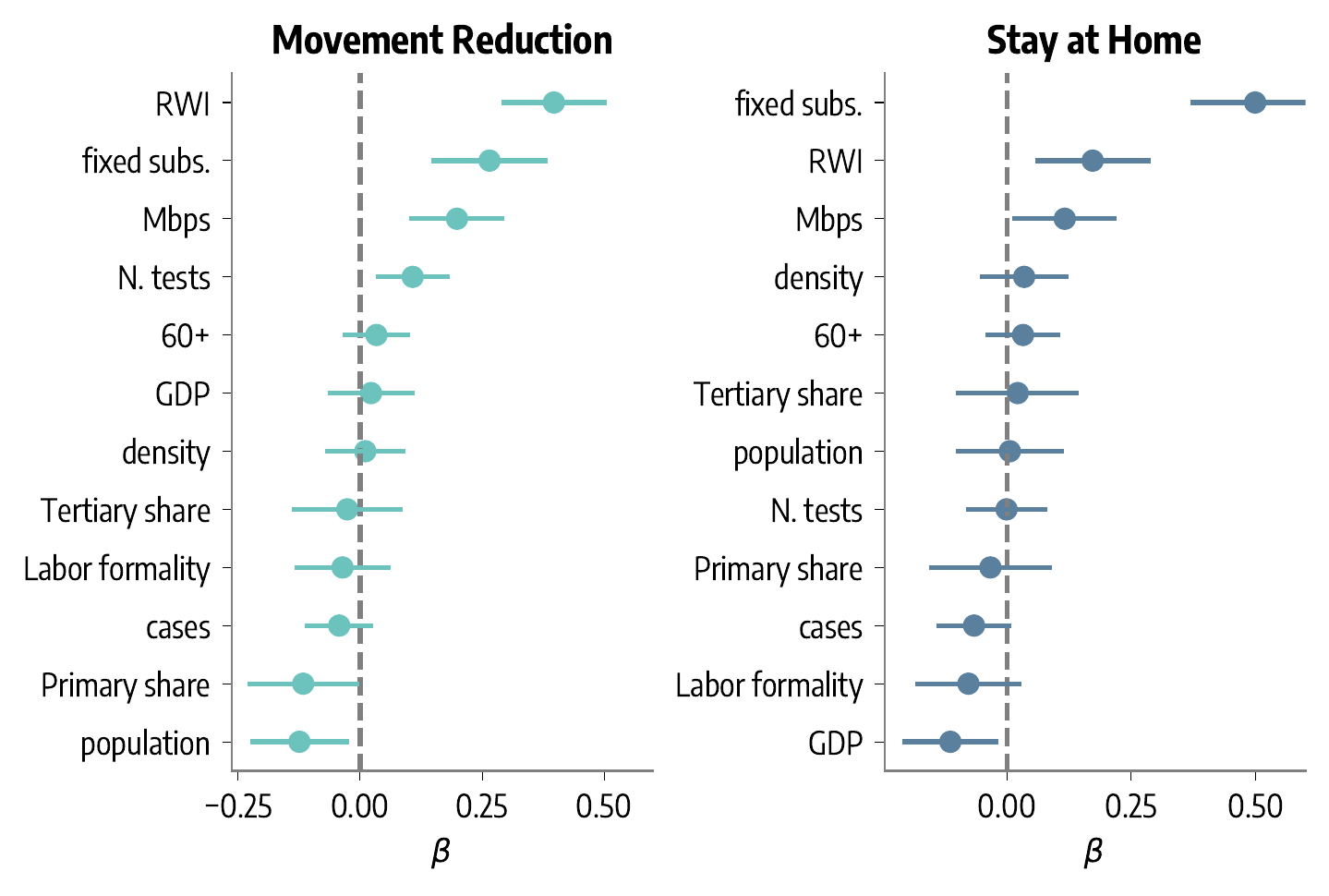}
    \caption{\textbf{Coefficients of the regressions for Colombia (employment data and importance of economic sectors included).}}
    \label{fig:coeffs_employment}
\end{figure}

We repeat the multiple variable regression including these additional variables. Coefficients are reported in Fig.~\ref{fig:coeffs_employment}. We notice that, even after including these additional features, the coefficients associated with digital infrastructure (\textit{Mbps}) and digital tools adoption (\textit{N. tests} and \textit{fixed subs}) are still positive and significant. The only exception is the number of consumer-initiated tests taken with Speedtest\textsuperscript{\tiny\textregistered} performed, whose coefficient is not significant and close to $0$ in the case of the \textit{stay at home} metric as dependent feature. Labor formality and the share of GDP associated with the tertiary sector is not significant in both cases. The coefficient of the share of GDP associated with the primary sector is instead negative and significant for movement reductions. This is in line with what expected, indeed the primary sector is mainly associated with essential activities that must continue also during lockdowns and that can hardly be performed remotely. 

\clearpage
\bibliography{ref}

\providecommand{\noopsort}[1]{}\providecommand{\singleletter}[1]{#1}%
\begin{thebibliography}{10}

\bibitem{perra2020non}
Nicola Perra.
\newblock {Non-pharmaceutical interventions during the COVID-19 pandemic: A
  review}.
\newblock {\em Physics Reports}, 2021.

\bibitem{Flaxman2020}
Seth Flaxman, Swapnil Mishra, Axel Gandy, H~Juliette~T Unwin, Thomas~A Mellan,
  Helen Coupland, Charles Whittaker, Harrison Zhu, Tresnia Berah, Jeffrey~W
  Eaton, M{\'{e}}lodie Monod, Pablo~N Perez-Guzman, Nora Schmit, Lucia Cilloni,
  Kylie E~C Ainslie, Marc Baguelin, Adhiratha Boonyasiri, Olivia Boyd, Lorenzo
  Cattarino, Laura~V Cooper, Zulma Cucunub{\'{a}}, Gina Cuomo-Dannenburg, Amy
  Dighe, Bimandra Djaafara, Ilaria Dorigatti, Sabine~L van Elsland, Richard~G
  FitzJohn, Katy A~M Gaythorpe, Lily Geidelberg, Nicholas~C Grassly, William~D
  Green, Timothy Hallett, Arran Hamlet, Wes Hinsley, Ben Jeffrey, Edward Knock,
  Daniel~J Laydon, Gemma Nedjati-Gilani, Pierre Nouvellet, Kris~V Parag, Igor
  Siveroni, Hayley~A Thompson, Robert Verity, Erik Volz, Caroline~E Walters,
  Haowei Wang, Yuanrong Wang, Oliver~J Watson, Peter Winskill, Xiaoyue Xi,
  Patrick G~T Walker, Azra~C Ghani, Christl~A Donnelly, Steven Riley, Michaela
  A~C Vollmer, Neil~M Ferguson, Lucy~C Okell, Samir Bhatt, and Imperial College
  COVID-19~Response Team.
\newblock {Estimating the effects of non-pharmaceutical interventions on
  COVID-19 in Europe}.
\newblock {\em Nature}, 584(7820):257--261, 2020.

\bibitem{Snoeijer2021}
Berber~T Snoeijer, Mariska Burger, Shaoxiong Sun, Richard J~B Dobson, and
  Amos~A Folarin.
\newblock {Measuring the effect of Non-Pharmaceutical Interventions (NPIs) on
  mobility during the COVID-19 pandemic using global mobility data}.
\newblock {\em npj Digital Medicine}, 4(1):81, 2021.

\bibitem{haug2020ranking}
Nils Haug, Lukas Geyrhofer, Alessandro Londei, Elma Dervic, Amelie
  Desvars-Larrive, Vittorio Loreto, Beate Pinior, Stefan Thurner, and Peter
  Klimek.
\newblock Ranking the effectiveness of worldwide covid-19 government
  interventions.
\newblock {\em Nature Human Behavior}, 2020.

\bibitem{cowling2020impact}
Benjamin~J Cowling, Sheikh~Taslim Ali, Tiffany~WY Ng, Tim~K Tsang, Julian~CM
  Li, Min~Whui Fong, Qiuyan Liao, Mike~YW Kwan, So~Lun Lee, Susan~S Chiu,
  et~al.
\newblock Impact assessment of non-pharmaceutical interventions against
  coronavirus disease 2019 and influenza in hong kong: an observational study.
\newblock {\em The Lancet Public Health}, 2020.

\bibitem{bonaccorsi2020economic}
Giovanni Bonaccorsi, Francesco Pierri, Matteo Cinelli, Andrea Flori, Alessandro
  Galeazzi, Francesco Porcelli, Ana~Lucia Schmidt, Carlo~Michele Valensise,
  Antonio Scala, Walter Quattrociocchi, et~al.
\newblock Economic and social consequences of human mobility restrictions under
  {COVID-19}.
\newblock {\em Proceedings of the National Academy of Sciences},
  117(27):15530--15535, 2020.

\bibitem{Skarp2021}
Janetta~E Skarp, Laura~E Downey, Julius W~E Ohrnberger, Lucia Cilloni,
  Alexandra~B Hogan, Abagael~L Sykes, Susannah~S Wang, Hiral~Anil Shah, Mimi
  Xiao, and Katharina Hauck.
\newblock {A Systematic Review of the Costs Relating to Non-pharmaceutical
  Interventions Against Infectious Disease Outbreaks}.
\newblock {\em Applied Health Economics and Health Policy}, 2021.

\bibitem{hbm1}
Irwin~M. Rosenstock.
\newblock The health belief model and preventive health behavior.
\newblock {\em Health Education Monographs}, 2(4):354--386, 1974.

\bibitem{hbm2}
G.M. Hochbaum.
\newblock {\em Public Participation in Medical Screening Programs: A
  Socio-psychological Study}.
\newblock Public Health Service publication. U.S. Department of Health,
  Education, and Welfare, Public Health Service, Bureau of State Services,
  Division of Special Health Services, Tuberculosis Program, 1958.

\bibitem{hbm3}
J.~Hayden.
\newblock {\em Introduction to Health Behavior Theory}.
\newblock Jones \& Bartlett Learning, 2013.

\bibitem{pullano2020population}
Giulia Pullano, Eugenio Valdano, Nicola Scarpa, Stefania Rubrichi, and Vittoria
  Colizza.
\newblock Population mobility reductions during {COVID-19} epidemic in {France}
  under lockdown.
\newblock {\em The Lancet Digital Health}, 2020.

\bibitem{Duenas2021}
Marco Due{\~{n}}as, Mercedes Campi, and Luis~E Olmos.
\newblock {Changes in mobility and socioeconomic conditions during the COVID-19
  outbreak}.
\newblock {\em Humanities and Social Sciences Communications}, 8(1):101, 2021.

\bibitem{Gozzi2021santiago}
Nicol{\`{o}} Gozzi, Michele Tizzoni, Matteo Chinazzi, Leo Ferres, Alessandro
  Vespignani, and Nicola Perra.
\newblock {Estimating the effect of social inequalities on the mitigation of
  COVID-19 across communities in Santiago de Chile}.
\newblock {\em Nature Communications}, 12(1):2429, 2021.

\bibitem{fraiberger2020uncovering}
Samuel~P Fraiberger, Pablo Astudillo, Lorenzo Candeago, Alex Chunet,
  Nicholas~KW Jones, Maham~Faisal Khan, Bruno Lepri, Nancy~Lozano Gracia,
  Lorenzo Lucchini, Emanuele Massaro, et~al.
\newblock Uncovering socioeconomic gaps in mobility reduction during the
  {COVID-19} pandemic using location data.
\newblock 2020.

\bibitem{Menaeabg5298}
Gonzalo~E. Mena, Pamela~P. Martinez, Ayesha~S. Mahmud, Pablo~A. Marquet,
  Caroline~O. Buckee, and Mauricio Santillana.
\newblock Socioeconomic status determines covid-19 incidence and related
  mortality in santiago, chile.
\newblock {\em Science}, 372(6545), 2021.

\bibitem{Chang2021}
Serina Chang, Emma Pierson, Pang~Wei Koh, Jaline Gerardin, Beth Redbird, David
  Grusky, and Jure Leskovec.
\newblock {Mobility network models of COVID-19 explain inequities and inform
  reopening}.
\newblock {\em Nature}, 589(7840):82--87, 2021.

\bibitem{topriceanu2020inequality}
Constantin-Cristian Topriceanu, Andrew Wong, James~C Moon, Alun Hughes, David
  Bann, Nishi Chaturvedi, Praveetha Patalay, Gabriella Conti, and Gabriella
  Captur.
\newblock Inequality in access to health and care services during
  lockdown-findings from the covid-19 survey in five uk national longitudinal
  studies.
\newblock {\em medRxiv}, 2020.

\bibitem{munoz2020racial}
L~Silvia Mu{\~n}oz-Price, Ann~B Nattinger, Frida Rivera, Ryan Hanson, Cameron~G
  Gmehlin, Adriana Perez, Siddhartha Singh, Blake~W Buchan, Nathan~A Ledeboer,
  and Liliana~E Pezzin.
\newblock Racial disparities in incidence and outcomes among patients with
  covid-19.
\newblock {\em JAMA Network Open}, 3(9):e2021892--e2021892, 2020.

\bibitem{yi2020health}
Huso Yi, Shu~Tian Ng, Aysha Farwin, Pei Ting~Amanda Low, Cheng~Mun Chang, and
  Jeremy Lim.
\newblock Health equity considerations in covid-19: geospatial network analysis
  of the covid-19 outbreak in the migrant population in singapore.
\newblock {\em Journal of Travel Medicine}, 2020.

\bibitem{mathur2020ethnic}
Rohini Mathur, Christopher~T Rentsch, Caroline Morton, William~J Hulme, Anna
  Schultze, Brian MacKenna, Rosalind~M Eggo, Krishnan Bhaskaran, Angel~YS Wong,
  Elizabeth~J Williamson, et~al.
\newblock Ethnic differences in covid-19 infection, hospitalisation, and
  mortality: an opensafely analysis of 17 million adults in england.
\newblock {\em medRxiv}, 2020.

\bibitem{blundell2020covid}
Richard Blundell, Monica Costa~Dias, Robert Joyce, and Xiaowei Xu.
\newblock Covid-19 and inequalities.
\newblock {\em Fiscal Studies}, 41(2):291--319, 2020.

\bibitem{Ahmed2020}
Faheem Ahmed, Na'eem Ahmed, Christopher Pissarides, and Joseph Stiglitz.
\newblock {Why inequality could spread COVID-19}.
\newblock {\em The Lancet Public Health}, 5(5):e240, may 2020.

\bibitem{strusani2020}
Davide Strusani and Georges~V. Houngbonon.
\newblock {What COVID-19 Means for Digital Infrastructure in Emerging Markets}.
\newblock {\em SSRN}, 2020.

\bibitem{covid19_digitalinfra}
Elisa~V. Mariscal, Alexander Elbittar, Martin Cave, Ruben Guerrero, Antonio
  Garcia-Zaballos, Enrique Iglesias, and William Webb.
\newblock {The Impact of Digital Infrastructure on the Consequences of COVID-19
  and on the Mitigation of Future Effects}.
\newblock {\em SSRN}, 2020.

\bibitem{work_digitaldivide}
Joseph Taylor and Rickey Taylor.
\newblock {Decreasing work-related movement during a pandemic. Location
  analytics and the implications of the digital divide}.
\newblock {\em International Journal of Development Issues}, 20:293--308, 2021.

\bibitem{Soomro2020}
Kamal~Ahmed Soomro, Ugur Kale, Reagan Curtis, Mete Akcaoglu, and Malayna
  Bernstein.
\newblock {Digital divide among higher education faculty}.
\newblock {\em International Journal of Educational Technology in Higher
  Education}, 17(1):21, 2020.

\bibitem{AZUBUIKE2021100022}
Obiageri~Bridget Azubuike, Oyindamola Adegboye, and Habeeb Quadri.
\newblock Who gets to learn in a pandemic? exploring the digital divide in
  remote learning during the covid-19 pandemic in nigeria.
\newblock {\em International Journal of Educational Research Open}, 2-2:100022,
  2021.

\bibitem{Eruchalu2021}
Chukwuma~N Eruchalu, Margaret~S Pichardo, Maheetha Bharadwaj, Carmen~B
  Rodriguez, Jorge~A Rodriguez, Regan~W Bergmark, David~W Bates, and Gezzer
  Ortega.
\newblock {The Expanding Digital Divide: Digital Health Access Inequities
  during the COVID-19 Pandemic in New York City}.
\newblock {\em Journal of Urban Health}, 98(2):183--186, 2021.

\bibitem{Watts2020}
Geoff Watts.
\newblock {COVID-19 and the digital divide in the UK}.
\newblock {\em The Lancet Digital Health}, 2(8):e395--e396, aug 2020.

\bibitem{vakataki2021visualizing}
Siope Vakataki‘Ofa and Cristina~Bernal Aparicio.
\newblock Visualizing broadband speeds in asia and the pacific.
\newblock 2021.

\bibitem{NBERw26982}
Lesley Chiou and Catherine Tucker.
\newblock Social distancing, internet access and inequality.
\newblock Working Paper 26982, National Bureau of Economic Research, April
  2020.

\bibitem{zhao2020effects}
Ying Zhao, Yong Guo, Yu~Xiao, Ranke Zhu, Wei Sun, Weiyong Huang, Deyi Liang,
  Liuying Tang, Fan Zhang, Dongsheng Zhu, et~al.
\newblock The effects of online homeschooling on children, parents, and
  teachers of grades 1--9 during the covid-19 pandemic.
\newblock {\em Medical Science Monitor: International Medical Journal of
  Experimental and Clinical Research}, 26:e925591--1, 2020.

\bibitem{elsalem2020stress}
Lina Elsalem, Nosayba Al-Azzam, Ahmad~A Jum'ah, Nail Obeidat, Amer~Mahmoud
  Sindiani, and Khalid~A Kheirallah.
\newblock Stress and behavioral changes with remote e-exams during the covid-19
  pandemic: A cross-sectional study among undergraduates of medical sciences.
\newblock {\em Annals of Medicine and Surgery}, 60:271--279, 2020.

\bibitem{torres2020transition}
Anna Torres, Ewa Doma{\'n}ska-Glonek, Wojciech Dzikowski, Jan Korulczyk, and
  Kamil Torres.
\newblock Transition to on-line is possible: solution for simulation-based
  teaching during pandemic.
\newblock {\em Medical education}, 2020.

\bibitem{passanisi2020quarantine}
Stefano Passanisi, Maria Pecoraro, Francesco Pira, Angela Alibrandi, Vittoria
  Donia, Paola Lonia, Giovanni~Battista Pajno, Giuseppina Salzano, and
  Fortunato Lombardo.
\newblock Quarantine due to the covid-19 pandemic from the perspective of
  pediatric patients with type 1 diabetes: a web-based survey.
\newblock {\em Frontiers in pediatrics}, 8:491, 2020.

\bibitem{jeste2020changes}
S~Jeste, C~Hyde, C~Distefano, A~Halladay, S~Ray, M~Porath, RB~Wilson, and
  A~Thurm.
\newblock Changes in access to educational and healthcare services for
  individuals with intellectual and developmental disabilities during covid-19
  restrictions.
\newblock {\em Journal of Intellectual Disability Research}, 64(11):825--833,
  2020.

\bibitem{Bauer2020OvercomingMH}
Johannes~M. Bauer, Keith~N. Hampton, Laleah Fernandez, and Craig~T. Robertson.
\newblock Overcoming michigan’s homework gap: The role of broadband internet
  connectivity for student success and career outlooks.
\newblock {\em IRPN: Innovation \& Information Management (Topic)}, 2020.

\bibitem{ookla}
{Ookla for Good}.
\newblock \url{https://www.speedtest.net/insights/blog/tag/ookla-for-good/},
  2021.
\newblock Accessed: 2021-09-07.

\bibitem{rangemaps}
{Movement Range Maps}.
\newblock \url{https://data.humdata.org/dataset/movement-range-maps?}, 2021.
\newblock Accessed: 2021-09-04.

\bibitem{lac_inequality}
Regional human development report for latin america and the caribbean 2010.
  acting on the future: Breaking the intergenerational transmission of
  inequality.
\newblock
  \url{https://www.latinamerica.undp.org/content/rblac/en/home/library/human_development/human-development-report.html},
  2010.
\newblock Accessed: 2021-09-06.

\bibitem{Lancet2020}
The Lancet.
\newblock {COVID-19 in Latin America: a humanitarian crisis}.
\newblock {\em The Lancet}, 396(10261):1463, nov 2020.

\bibitem{Lancet2021}
The Lancet.
\newblock {COVID-19 in Latin America: emergency and opportunity}.
\newblock {\em The Lancet}, 398(10295):93, jul 2021.

\bibitem{internet_users_lac}
{Individuals using the Internet (\% of population) - Latin America \&
  Caribbean}.
\newblock
  \url{https://data.worldbank.org/indicator/IT.NET.USER.ZS?locations=ZJ}.
\newblock Accessed: 2021-09-06.

\bibitem{wbinternetsubs}
{The World Bank, Fixed Broadband Subscriptions}.
\newblock \url{https://data.worldbank.org/indicator/IT.NET.BBND}, 2021.
\newblock Accessed: 2021-11-08.

\bibitem{Hale2021}
Thomas Hale, Noam Angrist, Rafael Goldszmidt, Beatriz Kira, Anna Petherick,
  Toby Phillips, Samuel Webster, Emily Cameron-Blake, Laura Hallas, Saptarshi
  Majumdar, and Helen Tatlow.
\newblock {A global panel database of pandemic policies (Oxford COVID-19
  Government Response Tracker)}.
\newblock {\em Nature Human Behaviour}, 5(4):529--538, 2021.

\bibitem{feamster2020measuring}
Nick Feamster and Jason Livingood.
\newblock Measuring internet speed: current challenges and future
  recommendations.
\newblock {\em Communications of the ACM}, 63(12):72--80, 2020.

\bibitem{ford2021form}
George~S Ford.
\newblock Form 477, speed-tests, and the american broadband user’s
  experience.
\newblock {\em Speed-Tests, and the American Broadband User’s Experience
  (March 31, 2021)}, 2021.

\bibitem{MORARIVERA2021102076}
Jorge Mora-Rivera and Fernando García-Mora.
\newblock Internet access and poverty reduction: Evidence from rural and urban
  mexico.
\newblock {\em Telecommunications Policy}, 45(2):102076, 2021.

\bibitem{MEDEIROS2021105118}
Victor Medeiros, Rafael Saulo~Marques Ribeiro, and Pedro Vasconcelos~Maia
  do~Amaral.
\newblock Infrastructure and household poverty in brazil: A regional approach
  using multilevel models.
\newblock {\em World Development}, 137:105118, 2021.

\bibitem{galperin2017internet_poverty}
Hernan Galperin and M.~Fernanda~Viecens.
\newblock Connected for development? theory and evidence about the impact of
  internet technologies on poverty alleviation.
\newblock {\em Development Policy Review}, 35(3):315--336, 2017.

\bibitem{Kim2015}
Seongho Kim.
\newblock {ppcor: An R Package for a Fast Calculation to Semi-partial
  Correlation Coefficients}.
\newblock {\em Communications for statistical applications and methods},
  22(6):665--674, nov 2015.

\bibitem{Hair.2009}
J.~Hair, R.~Anderson, and B.~Babin.
\newblock {\em {Multivariate Data Analysis}}.
\newblock Prentice Hall, 7 edition, 2009.

\bibitem{doi:10.1098/rsif.2020.1035}
Samuel Heroy, Isabella Loaiza, Alex Pentland, and Neave O’Clery.
\newblock Covid-19 policy analysis: labour structure dictates lockdown mobility
  behaviour.
\newblock {\em Journal of The Royal Society Interface}, 18(176):20201035, 2021.

\bibitem{LEE2021102563}
Won~Do Lee, Matthias Qian, and Tim Schwanen.
\newblock The association between socioeconomic status and mobility reductions
  in the early stage of england's covid-19 epidemic.
\newblock {\em Health \& Place}, 69:102563, 2021.

\bibitem{Chang2021_socioecon}
Hsien-Yen Chang, Wenze Tang, Elham Hatef, Christopher Kitchen, Jonathan~P
  Weiner, and Hadi Kharrazi.
\newblock {Differential impact of mitigation policies and socioeconomic status
  on COVID-19 prevalence and social distancing in the United States}.
\newblock {\em BMC Public Health}, 21(1):1140, 2021.

\bibitem{gauvin_socioecon}
Laetitia Gauvin, Paolo Bajardi, Emanuele Pepe, Brennan Lake, Filippo Privitera,
  and Michele Tizzoni.
\newblock Socio-economic determinants of mobility responses during the first
  wave of covid-19 in italy: from provinces to neighbourhoods.
\newblock {\em Journal of The Royal Society Interface}, 18(181):20210092, 2021.

\bibitem{akaike1998information}
Hirotogu Akaike.
\newblock Information theory and an extension of the maximum likelihood
  principle.
\newblock In {\em Selected papers of hirotugu akaike}, pages 199--213.
  Springer, 1998.

\bibitem{Portet2020}
St{\'{e}}phanie Portet.
\newblock {A primer on model selection using the Akaike Information Criterion}.
\newblock {\em Infectious Disease Modelling}, 5:111--128, jan 2020.

\bibitem{info:doi/10.2196/19361}
Allison Crawford and Eva Serhal.
\newblock Digital health equity and covid-19: The innovation curve cannot
  reinforce the social gradient of health.
\newblock {\em J Med Internet Res}, 22(6):e19361, Jun 2020.

\bibitem{doi:10.1098/rsif.2012.0986}
Amy Wesolowski, Nathan Eagle, Abdisalan~M. Noor, Robert~W. Snow, and
  Caroline~O. Buckee.
\newblock {The impact of biases in mobile phone ownership on estimates of human
  mobility}.
\newblock {\em Journal of The Royal Society Interface}, 10(81):20120986, 2013.

\bibitem{10.1371/journal.pcbi.1003716}
Michele Tizzoni, Paolo Bajardi, Adeline Decuyper, Guillaume Kon Kam~King,
  Christian~M. Schneider, Vincent Blondel, Zbigniew Smoreda, Marta~C.
  González, and Vittoria Colizza.
\newblock {On the Use of Human Mobility Proxies for Modeling Epidemics}.
\newblock {\em PLOS Computational Biology}, 10(7):1--15, 07 2014.

\bibitem{10.1145/3442188.3445881}
Amanda Coston, Neel Guha, Derek Ouyang, Lisa Lu, Alexandra Chouldechova, and
  Daniel~E. Ho.
\newblock {Leveraging Administrative Data for Bias Audits: Assessing Disparate
  Coverage with Mobility Data for COVID-19 Policy}.
\newblock In {\em Proceedings of the 2021 ACM Conference on Fairness,
  Accountability, and Transparency}, FAccT '21, page 173–184, New York, NY,
  USA, 2021. Association for Computing Machinery.

\bibitem{schlosser2021biases}
Frank Schlosser, Vedran Sekara, Dirk Brockmann, and Manuel Garcia-Herranz.
\newblock Biases in human mobility data impact epidemic modeling, 2021.

\bibitem{oxford_codebook}
{Variation in government responses to COVID-19}.
\newblock
  \url{https://www.bsg.ox.ac.uk/sites/default/files/2021-06/BSG-WP-2020-032-v12_0.pdf},
  2021.
\newblock Accessed: 2021-09-04.

\bibitem{Kishore2021}
Nishant Kishore, Rebecca Kahn, Pamela~P Martinez, Pablo~M {De Salazar},
  Ayesha~S Mahmud, and Caroline~O Buckee.
\newblock {Lockdowns result in changes in human mobility which may impact the
  epidemiologic dynamics of SARS-CoV-2}.
\newblock {\em Scientific Reports}, 11(1):6995, 2021.

\bibitem{Cortes2021}
Ulises Cort{\'{e}}s, Atia Cort{\'{e}}s, Dario Garcia-Gasulla, Raquel
  P{\'{e}}rez-Arnal, Sergio {\'{A}}lvarez-Napagao, and Enric {\`{A}}lvarez.
\newblock {The ethical use of high-performance computing and artificial
  intelligence: fighting COVID-19 at Barcelona Supercomputing Center}.
\newblock {\em AI and Ethics}, 2021.

\bibitem{rangemaps_methods}
{Protecting privacy in Facebook mobility data during the COVID-19 response}.
\newblock
  \url{https://research.fb.com/blog/2020/06/protecting-privacy-in-facebook-mobility-data-during-the-covid-19-response/},
  2021.
\newblock Accessed: 2021-09-04.

\bibitem{chi2021microestimates}
Guanghua Chi, Han Fang, Sourav Chatterjee, and Joshua~E. Blumenstock.
\newblock {Micro-Estimates of Wealth for all Low- and Middle-Income Countries},
  2021.

\bibitem{rwi_calculation}
{Tutorial: Calculating Population Weighted Relative Wealth Index}.
\newblock
  \url{https://dataforgood.facebook.com/dfg/docs/tutorial-calculating-population-weighted-relative-wealth-index},
  2021.
\newblock Accessed: 2021-11-19.

\bibitem{popDANE}
{National Administrative Department of Statistics of Colombia - Population
  Estimates}.
\newblock
  \url{https://www.dane.gov.co/index.php/estadisticas-por-tema/demografia-y-poblacion/proyecciones-de-poblacion},
  2021.
\newblock Accessed: 2021-11-19.

\bibitem{density_maps}
{High Resolution Population Density Maps and Demographic Estimates}.
\newblock
  \url{https://dataforgood.fb.com/docs/high-resolution-population-density-maps-demographic-estimates-documentation/},
  2021.
\newblock Accessed: 2021-09-07.

\bibitem{colombia_covid19cases}
{Casos positivos de COVID-19 en Colombia}.
\newblock
  \url{https://www.datos.gov.co/Salud-y-Protecci-n-Social/Casos-positivos-de-COVID-19-en-Colombia/gt2j-8ykr/data},
  2021.
\newblock Accessed: 2021-09-07.

\bibitem{covid_cases_ecuador}
{Ecuador, Evolución del coronavirus por cantones}.
\newblock \url{https://www.covid19ecuador.org/cantones}, 2021.
\newblock Accessed: 2022-01-04.

\bibitem{elsalvador_covid19cases}
{Datos diarios de COVID 19 en El Salvador}.
\newblock \url{https://diario.innovacion.gob.sv}, 2021.
\newblock Accessed: 2021-09-07.

\bibitem{dane_gdp}
{National Administrative Department of Statistics of Colombia, Cuentas
  nacionales departamentales: PIB por departamento}.
\newblock
  \url{https://www.dane.gov.co/index.php/estadisticas-por-tema/cuentas-nacionales/cuentas-nacionales-departamentales},
  2021.
\newblock Accessed: 2022-01-04.

\bibitem{hdi_ecuador}
J.~Illingworth and F.~Campaña.
\newblock {\em {Informe sobre Desarrollo Humano del Ecuador}}.
\newblock Fundación Ecuador, 2019.

\bibitem{almanaque262}
{Almanaque 262. Estado del desarrollo humano en los municipios de El Salvador}.
\newblock
  \url{https://www.sv.undp.org/content/el_salvador/es/home/library/hiv_aids/almanaque-262.html},
  2021.
\newblock Accessed: 2021-09-07.

\bibitem{gadm}
{Database of Global Administrative Areas}.
\newblock \url{https://gadm.org/download_country.html}, 2021.
\newblock Accessed: 2022-01-04.

\bibitem{dane_mpi}
{Medida de pobreza multidimensional municipal de fuente censal 2018}.
\newblock
  \url{https://www.dane.gov.co/index.php/estadisticas-por-tema/pobreza-y-condiciones-de-vida/pobreza-y-desigualdad/medida-de-pobreza-multidimensional-de-fuente-censal},
  2021.
\newblock Accessed: 2021-09-07.

\bibitem{GVK330798693}
{Richard Arnold} Johnson and {Dean W.} Wichern.
\newblock {\em Applied multivariate statistical analysis}.
\newblock Prentice Hall, 5. ed edition, 2002.

\bibitem{Vallat2018}
Raphael Vallat.
\newblock Pingouin: statistics in python.
\newblock {\em Journal of Open Source Software}, 3(31):1026, 2018.

\bibitem{10.2307/1271436}
Arthur~E. Hoerl and Robert~W. Kennard.
\newblock Ridge regression: Biased estimation for nonorthogonal problems.
\newblock {\em Technometrics}, 42(1):80--86, 2000.

\bibitem{scikit-learn}
F.~Pedregosa, G.~Varoquaux, A.~Gramfort, V.~Michel, B.~Thirion, O.~Grisel,
  M.~Blondel, P.~Prettenhofer, R.~Weiss, V.~Dubourg, J.~Vanderplas, A.~Passos,
  D.~Cournapeau, M.~Brucher, M.~Perrot, and E.~Duchesnay.
\newblock Scikit-learn: Machine learning in {P}ython.
\newblock {\em Journal of Machine Learning Research}, 12:2825--2830, 2011.

\bibitem{daneinternet}
{Colombia - Fixed Internet Penetration at the Municipal Level}.
\newblock
  \url{https://www.datos.gov.co/Ciencia-Tecnolog-a-e-Innovaci-n/Internet-Fijo-Penetraci-n-Municipio/fut2-keu8},
  2021.
\newblock Accessed: 2021-11-08.

\bibitem{o2016path}
Neave O’Clery, Andres Gomez-Lievano, and Eduardo Lora.
\newblock The path to labor formality: urban agglomeration and the emergence of
  complex industries.
\newblock {\em CID Research Fellow and Graduate Student Working Paper Series},
  2016.

\bibitem{datlascolombia}
{El Atlas Colombiano de Complejidad Económica}.
\newblock \url{http://datlascolombia.com/#/downloads}, 2021.
\newblock Accessed: 2022-01-04.

\bibitem{colombia_pop_hdx}
{Colombia - Subnational Population Statistics}.
\newblock
  \url{https://data.humdata.org/dataset/colombia-population-estimates-and-projections},
  2021.
\newblock Accessed: 2022-01-04.

\end{thebibliography}

\end{document}